\documentclass[journal]{IEEEtran}
\usepackage{amsmath, amssymb, cite, tikz, graphicx, mathrsfs, mathtools, psfrag, subfigure,color,soul,graphicx}
\usepackage{setspace,float,pstool}
\usepackage{relsize}
\usepackage{enumitem}
\usepackage{bbm}
\usepackage[utf8]{inputenc}
\usepackage{booktabs}
\usepackage{amscd}
\usepackage{amsmath}
\usepackage{amssymb}
\usepackage{amsthm}

\usepackage{epsfig}
\usepackage{verbatim}
\usepackage{graphicx}
\usepackage{amsthm}
\usepackage{color}
\usepackage{ multirow }
\usepackage{algorithm}
\usepackage[noend]{algpseudocode}
\usetikzlibrary{patterns}
\usetikzlibrary{angles} 
\usepackage{tikz}
\usetikzlibrary{shapes, arrows, decorations.markings, arrows.meta}
\usetikzlibrary{patterns} 
\usetikzlibrary{arrows,automata,positioning,chains}
\usetikzlibrary{quotes,angles}
\newtheorem{remark}{Remark}
\def\BState{\State\hskip-\ALG@thistlm}
\makeatother
\newtheorem{theorem}{Theorem}
\newtheorem{lemma}[theorem]{Lemma}
\begin{document}

\title{Real-Time Remote Monitoring of Correlated Markovian Sources}
\author{Mehrdad Salimnejad, Marios Kountouris, \IEEEmembership{Fellow, IEEE}, and Nikolaos Pappas, 
\IEEEmembership{Senior Member, IEEE}.
\thanks{M. Salimnejad and N. Pappas are with the Department of Computer and Information Science Linköping University, Sweden, email: \{\texttt{mehrdad.salimnejad, nikolaos.pappas\}@liu.se}. M. Kountouris is with the Department of Computer Science and Artificial Intelligence, Andalusian Research Institute in Data Science and Computational Intelligence (DaSCI), University of Granada, Spain, email: \texttt{mariosk@ugr.es}.\\ The work of M. Salimnejad and N. Pappas has been supported by the Swedish Research Council (VR), ELLIIT, and the European Union (6G-LEADER, 101192080, SOVEREIGN, 101131481, and ROBUST-6G, 101139068). The work of M. Kountouris has been supported by the European Research Council (ERC) under the European Union’s Horizon 2020 research and innovation programme (Grant agreement No. 101003431).}}

\maketitle

\begin{abstract}
We investigate real-time tracking of two correlated stochastic processes over a shared wireless channel. The joint evolution of the processes is modeled as a two-dimensional discrete-time Markov chain. Each process is observed by a dedicated sampler and independently reconstructed at a remote monitor according to a task-specific objective. Although both processes originate from a common underlying phenomenon (e.g., distinct features of the same source), each monitor is interested only in its corresponding feature. A reconstruction error is incurred when the true and reconstructed states mismatch at one or both monitors. To address this problem, we propose an \emph{error-aware} joint sampling and transmission policy, under which each sampler probabilistically generates samples only when the current process state differs from the most recently reconstructed state at its corresponding monitor. We adopt the time-averaged reconstruction error as the primary performance metric and benchmark the proposed policy against state-of-the-art joint sampling and transmission schemes. For each policy, we derive closed-form expressions for the resulting time-averaged reconstruction error. We further formulate and solve an optimization problem that minimizes the time-averaged reconstruction error subject to an average sampling cost constraint. Analytical and numerical results demonstrate that the proposed error-aware policy achieves the minimum time-averaged reconstruction error among the considered schemes while efficiently utilizing the sampling budget. The performance gains are particularly pronounced in regimes with strong inter-process correlation and stringent tracking requirements, where frequent sampling by both samplers is necessary.
\end{abstract}


\maketitle

\section{INTRODUCTION}
Real-time communication systems are fundamental to time-sensitive applications such as autonomous transportation, swarm robotics, industrial control, and healthcare monitoring~\cite{hult2016coordination, park2018wireless, pappas2022agebook,shreedhar2019age, vitturi2013industrial}. In these systems, sensors or intelligent agents continuously observe dynamic processes and convey status updates to remote monitors over bandwidth-limited and unreliable wireless links. The received updates are then processed to extract relevant and actionable information that enables timely decision-making and control. Because the performance of such applications critically depends on the \emph{freshness, relevance, and significance} of the delivered information, it is imperative to design resource-efficient mechanisms for data generation, transmission, and utilization. However, stringent latency, energy, and bandwidth constraints often preclude transmitting all observations, necessitating principled policies that decide which updates to generate and when to communicate them. This motivation has led to \emph{goal-oriented, semantics-aware communication}, a novel paradigm for status update systems that prioritizes delivering the right information at the right time to achieve system objectives~\cite{kountouris2021semantics,kalfa2021towards,popovski2020semantic,popovski2022perspective,gunduz2022beyond, utkovski2023semantic, luo2025informationfreshnesssemanticsinformation}. Central to this framework are semantics-aware metrics that quantify the timeliness and utility of information, including Age of Information (AoI) \cite{kaul2012real}, Age of Incorrect Information (AoII) \cite{maatouk2020age}, cost of actuation error \cite{pappas2021goal}, and Version Age of Information (VAoI) \cite{yates2021age}. Prior work has developed sampling and transmission policies that optimize these metrics under diverse constraints, demonstrating substantial gains in the efficiency and effectiveness of real-time monitoring. In multi-source systems, however, the marginal value of an update is inherently \emph{conditional} on the receiver’s information state, which can already contain partial knowledge about other sources due to statistical correlation. However, most existing results focus on a single information source generating updates independently (e.g., as an independent and identically distributed (i.i.d.) process or a Markov chain), and therefore do not capture multi-source settings with correlated observations. In many practical deployments, such as multi-robot mapping with overlapping fields of view or dense wireless sensor networks, updates from one source can be informative about the states of others, implying that correlation should be explicitly accounted for when designing semantics-aware sampling and scheduling policies.

In this paper, we study a time-slotted communication system in which two samplers independently observe two correlated processes drawn from a common joint information source. After each sampling decision, a sampler transmits its update as a packet to its corresponding remote receiver over a shared and unreliable wireless channel. Each receiver reconstructs the state of its observed process using the packets received successfully. A reconstruction error occurs whenever the true process state differs from the receiver’s reconstructed state. We propose a joint sampling and transmission policy, termed the \emph{error-aware} policy, under which each sampler probabilistically chooses to sample when its current state differs from the receiver’s reconstructed state in the previous time slot. Our analysis explicitly leverages inter-source correlation to quantify the marginal value of an update in closed form, enabling principled policy design beyond the independent-source setting. To quantify performance, we derive a closed-form expression for the long-term time-averaged reconstruction error under the error-aware policy and benchmark it against representative state-of-the-art joint sampling and transmission strategies. The resulting characterization yields structural insights into the optimal sampling behavior, revealing when correlation and channel unreliability induce threshold-like operating regimes. We then formulate and solve an optimization problem that minimizes the time-averaged reconstruction error subject to a constraint on the time-averaged sampling cost. Finally, we characterize how key system parameters influence the optimal operating point via analytical insights and supporting numerical results.
\subsection{Related Work}
The primary focus of this study is to address real-time remote tracking of a correlated information source while explicitly accounting for the source dynamics relevant to actuation. A substantial body of prior work has investigated real-time remote monitoring and estimation. The works in \cite{xu2004optimal, shi2011sensor, shi2011time, wu2013can, li2010event, haakansson2021optimal} study scheduling for event-triggered estimation, where a sensor observes the source state and transmits updates only when prescribed events occur, aiming to minimize estimation error subject to communication-rate constraints. Complementary lines of research incorporate the structure and dynamics of the underlying processes in the design of monitoring policies~\cite{shi2012scheduling,wu2018optimal,chakravorty2015distortion,chakravorty2016fundamental,sun2019sampling,ornee2021sampling,guo2021optimal,hui2022real,chakravorty2019remote,Chen2017}. These works analyze optimal transmission and sampling strategies for processes such as Gauss–Markov models \cite{shi2012scheduling,wu2018optimal}, characterize fundamental trade-offs between communication resources and estimation accuracy \cite{chakravorty2015distortion,chakravorty2016fundamental}, stochastic processes \cite{sun2019sampling,ornee2021sampling,guo2021optimal,hui2022real}, and finite-state Markov chains \cite{chakravorty2019remote,Chen2017}. Overall, these studies primarily target minimizing estimation error through sampling and transmission design but do not explicitly account for the \emph{informativeness} of updates or their downstream impact on actuation and control. The works \cite{salimnejad2023state,salimnejad2024real,cocco2023remote,santi2024remote,luo2025semantic,talli2025pragmatic,salimnejad2024age,delfani2024semantics,JipingTIT2025,Li2024,Vilni2024,luo2025informationfreshnesssemanticsinformation} propose semantics-aware performance metrics that assess the \emph{effectiveness} of information by leveraging the synergies between data processing, information transmission, and signal reconstruction. However, all the aforementioned works do not study multi-source settings with correlated observations in the design of joint sampling and transmission policies or semantics-aware performance metrics.
The monitoring of correlated sources has  been investigated in~\cite{poojary2017real,jiang2019status,zhou2020age,kalor2019minimizing,tripathi2022optimizing,ramakanth2024monitoring}. These works address challenges such as temporal correlations \cite{poojary2017real}, spatial correlations \cite{jiang2019status}, sensor scheduling \cite{zhou2020age, kalor2019minimizing}, and the minimization of monitoring error or AoI \cite{tripathi2022optimizing, ramakanth2024monitoring}. In particular, \cite{poojary2017real, zhou2020age, jiang2019status, kalor2019minimizing} study strategies such as optimizing sensor placement, device scheduling, and transmission policies to reduce estimation error and energy consumption, while \cite{tripathi2022optimizing, ramakanth2024monitoring} propose frameworks for monitoring multiple correlated sources under observation constraints. Despite these advances, existing studies largely focus on timeliness- or error-based objectives (e.g., AoI or estimation/monitoring error) and do not explicitly incorporate the \emph{importance} or \emph{usefulness} of information for downstream system performance, particularly in actuation-oriented settings.

\subsection{Summary of Contributions}
Nevertheless, existing approaches do not jointly account for source correlation and actuation-oriented semantics in the design and analysis of sampling and transmission policies. In this paper, we address this gap. Specifically, we consider correlated sources communicating over a shared unreliable channel and develop analytically tractable joint sampling and transmission policies with performance guarantees under an actuation-relevant reconstruction-error metric. We investigate real-time remote monitoring of a correlated information source while explicitly accounting for the source dynamics relevant to actuation. The main contributions of this paper are as follows:
\begin{enumerate}
    \item We introduce a joint sampling and transmission policy, termed the \emph{error-aware} policy, in which each sampler probabilistically decides whether to generate a sample based on a comparison between its current process state and the receiver’s reconstructed state from the previous time slot. By skipping redundant samples when no state mismatch is present, the proposed policy is particularly well suited to resource-constrained environments, such as energy-limited wireless networks, where efficient use of sampling and transmission resources is critical.
    \item We derive closed-form expressions for the time-averaged reconstruction error achieved by the proposed error-aware policy. For comparison, we also obtain analytical expressions for the time-averaged reconstruction error under the semantics-aware, change-aware, and randomized stationary sampling-and-transmission policies proposed in~\cite{pappas2021goal,salimnejad2024real}.
    \item We formulate and solve a constrained optimization problem to determine the optimal sampling probabilities for both the randomized stationary and error-aware policies, with the objective of minimizing the time-averaged reconstruction error subject to a constraint on the time-averaged sampling cost. The resulting solutions demonstrate that the optimized error-aware policy consistently outperforms the considered baseline schemes in terms of reconstruction accuracy, while efficiently exploiting the available sampling budget.
\end{enumerate}
The remainder of the paper is organized as follows. Section~II describes the system model and introduces the considered sampling-and-transmission policies. Section~III defines the time-averaged reconstruction error and derives closed-form expressions for it under the considered policies. Section~IV formulates the sampling-cost–constrained optimization problem and discusses its solution. Section~V presents numerical results that validate the analysis and illustrate the impact of key system parameters. Section~VI concludes the paper. Technical proofs are provided in the Appendix.

\section{System Model}
\begin{figure*}[t!]
    \centering
    \includegraphics[width=1\linewidth]{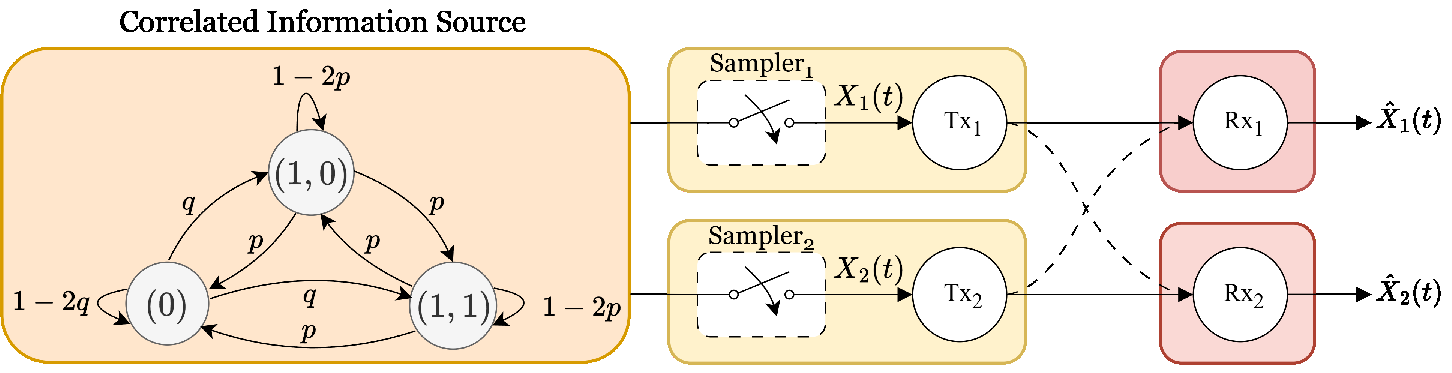}
 \vspace{-0.6cm}
    \caption{Real-time remote monitoring of a correlated information source over wireless.}
    \label{system_model_fig}
\end{figure*}

We consider a slotted-time communication system in which two samplers independently observe a joint information source, while two transmitters ($\text{Tx}_{1}$ and $\text{Tx}_{2}$) send their updates, in the form of packets, to two receivers ($\text{Rx}_{1}$ and $\text{Rx}_{2}$) in support of their respective monitoring goals, as illustrated in Fig.~\ref{system_model_fig}. The information source is modeled as a two-dimensional discrete-time Markov chain (DTMC) composed of two correlated processes, each observed by one sampler. Each process represents an attribute or feature that contributes to the overall state of the source. The DTMC captures the joint evolution of these attributes, thereby reflecting their temporal statistical dependence. 

Let $X_m(t)$ denote the state of the $m$-th process at time slot $t$, taking values in a finite set $\mathcal{X}_m = \{X_{m,i}\,|\,i=1, 2, \cdots, |\mathcal{X}_m|\}$. The processes $X_1(t)$ and $X_2(t)$ are statistically correlated, with joint distribution $P_{i,j} = \mathbb{P}\big[X_1(t)=X_{1,i}, X_2(t)=X_{2,j}\big]$, $\forall X_{1,i} \in \mathcal{X}_1$, $\forall X_{2,j} \in \mathcal{X}_2$. 
To keep the analysis tractable while preserving the essential insights of this work, we focus on a three-state DTMC, shown in Fig. \ref{system_model_fig}. The system state at time slot $t$ is represented by the tuple $\big(X_1(t), X_2(t)\big)$, and its evolution is governed by the transition matrix $Q = [P_{i,j/ i^\prime, j^\prime}]_{3\times 3}$ with the elements defined as $P_{i,j/ i^\prime, j^\prime} = \mathbb{P}\big[X_{1}(t+1)=X_{1,i^\prime},X_{2}(t+1)=X_{2,j^\prime}
   \big|X_{1}(t)=X_{1,i}, X_{2}(t)=X_{2,j}\big].$
Note that in the considered correlated information source model, as depicted in Fig. \ref{system_model_fig}, when the DTMC is in state $0$ at time slot $t$, only the first process carries information (i.e., $X_1(t) = 0$). In this case, if $\text{Rx}_{1}$ is synchronized at that slot, then both receivers are synchronized; otherwise, both are in error. Thus, updates from the first process implicitly convey information about the second process. With this consideration, we capture an inherent prioritization among the processes, since when $X_1(t) = 0$, the value of the second process is irrelevant.

At the beginning of each time slot $t$, each sampler observes the state of its corresponding process and decides whether to take a sample. The sampling action of the $m$-th sampler is denoted by $\alpha_{m}(t) \in \{0,1\}$, where $\alpha_{m}(t) = 1$ indicates that a sample is taken and $\alpha_{m}(t) = 0$ otherwise. If a sample is taken at time slot $t$, the resulting update packet is transmitted immediately within the same time slot over wireless channels that share a common frequency band. As a result, when both samplers are active simultaneously, the packets transmitted to $\text{Rx}_{1}$ and $\text{Rx}_{2}$ interfere with each other in addition to being affected by channel noise. In this system, each receiver is responsible for tracking one of the attribute processes according to its monitoring objective. Specifically, $\text{Rx}_{1}$ and $\text{Rx}_{2}$ monitor the processes sampled by $\text{Sampler}_1$ and $\text{Sampler}_2$, respectively. At the end of time slot $t$, the $m$-th receiver constructs an estimate of the state of its corresponding process, denoted by $\hat{X}_{m}(t) \in \mathcal{X}_m$, based on the most recently received successfully decoded update from its associated sampler. The reconstructed state $\hat{X}_{m}(t)$ is assumed to be decoded successfully with probability $p_{s_{m/m}}$ when only the $m$-th transmitter is active, and with probability $p_{s_{m/nm}}$ when both transmitters are simultaneously active, where $n \neq m \in \{1,2\}$. We assume that at time slot $t$, if the $m$-th process is not sampled or if the $m$-th Rx fails to decode the transmitted update, the reconstructed state remains unchanged, i.e., $\hat{X}_m(t) = \hat{X}_m(t-1)$. The $m$-th receiver is said to be \textit{synchronized} (sync) at time slot $t$ if $X_m(t) = \hat{X}_m(t)$, and in an \textit{erroneous state} if $X_m(t) \neq \hat{X}_m(t)$. Acknowledgment (ACK) and negative acknowledgment (NACK) packets are used to notify the transmitters of the success or failure of each transmission, and these feedback messages are assumed to be delivered instantaneously and without errors\footnote{An ACK/NACK feedback channel is required to implement the error-aware and semantics-aware policies.}. Therefore, at each time slot $t$, the transmitters have perfect knowledge of the reconstructed source states, i.e., $\hat{X}_{m}(t), \forall m \in \{1, 2\}$. In addition, we assume that any sample corresponding to an unsuccessful transmission is discarded (i.e., the system operates over a packet-drop channel). 

We next propose a joint sampling and transmission strategy, referred to as the \emph{error-aware} policy, in which each sampler triggers sampling probabilistically. Under this policy, the $m$-th sampler remains idle if the current state of its process matches the reconstructed state from the previous time slot; otherwise, it samples the process with probability $q_{\alpha_{m}}$. 

For comparison, we also consider three existing sampling policies, namely the randomized stationary, change-aware, and semantics-aware policies, introduced in \cite{pappas2021goal} and \cite{salimnejad2024real}. A brief description of each policy is provided below.
\begin{enumerate}
    \item Randomized Stationary (RS): At each time slot, a new sample is generated probabilistically, independent of the receiver's synchronization state. Specifically, the $m$-th sampler takes a sample with probability $p_{\alpha_{m}} = \mathbb{P}[\alpha_{m}(t) = 1]$ and remains idle with probability $1 - p_{\alpha_{m}}$.
    \item Change-aware (CA): The $m$-th sampler takes a new sample at time slot $t$ if and only if the state of its process differs from its state in the previous slot, i.e., if $X_{m}(t) \neq X_{m}(t-1)$, regardless of whether the $m$-th receiver is synchronized.
    \item Error-aware (EA): At time slot $t$, if the $m$-th receiver is in sync, the $m$-th sampler remains idle. Otherwise, it generates a new sample with probability $q_{\alpha_{m}} = \mathbb{P}[\alpha_{m}(t) = 1 \mid X_{m}(t) \neq \hat{X}_{m}(t-1)]$ and stays idle with probability $1 - q_{\alpha_{m}}$.
    \item Semantics-aware (SA): The $m$-th sampler generates a sample at time slot $t$ whenever the state of its process differs from the reconstructed state at the previous time slot, i.e., when $X_{m}(t) \neq \hat{X}_{m}(t-1)$.\footnote{The semantics-aware policy is a special case of the error-aware policy with $q_{\alpha_{m}} = 1, \forall m\in\{1,2\}$.}
\end{enumerate}
\section{Performance Metric and Analysis}
In this section, we analyze the impact of information semantics at the receivers. To this end, we evaluate the system performance using a key metric, namely the \emph{time-averaged reconstruction error}. This metric quantifies the discrepancy between the source's original state and its reconstructed state over time, thus capturing the effectiveness of the semantics-aware transmission strategy.
\vspace{-0.3cm}
\subsection{Time-averaged Reconstruction Error}
The performance of a real-time tracking system is determined by the actions or effects produced by the endpoint over a period of time. Since these actions rely on the accurate reconstruction of source attribute at both receivers, we introduce the time-averaged reconstruction error as a key performance metric. This error is modeled as a composite function $f$ that captures the discrepancy errors occurring at the receivers over time. Accordingly, the time-averaged reconstruction error $P_E$ over the observation interval $[1, T]$, where $T$ is a large positive integer, is defined as follows:
\vspace{-0.2cm}
\begin{align}
\label{Taveg_Error}
    P_{E}  = \underset{T \rightarrow \infty}{\lim} \frac{1}{T} \sum_{t=1}^{T} f\big(E_{1}(t),E_{2}(t)\big),
\end{align}
where $E_{m}(t)$ denotes the discrepancy error at the $m$-th receiver, for $m\in\{1,2\}$, at time slot $t$, is defined as:
\begin{align}
\label{Emt}
    E_{m}(t) = \big|X_m(t)- \hat{X}_m(t)\big|.
\end{align}
We now assume that the system is in an erroneous state whenever the sum of the discrepancy errors is nonzero. Accordingly, the function $f(\cdot, \cdot)$ in \eqref{Taveg_Error} is defined as follows:
\begin{align}
    \label{PE_f}
f\big(E_{1}(t),E_{2}(t)\big) = \mathbbm1\big(E_{1}(t)+E_{2}(t)\neq 0\big),
\end{align}
where $\mathbbm1(\cdot)$ denotes the indicator function. 
Substituting \eqref{PE_f} in \eqref{Taveg_Error}, the time-averaged reconstruction error becomes:
\begin{align}
\label{Taveg_Error2}
    P_{E}  = \underset{T \rightarrow \infty}{\lim} \frac{1}{T} \sum_{t=1}^{T} \mathbbm1\big(E_{1}(t)+E_{2}(t)\neq 0\big).
\end{align}

\begin{lemma}
		\label{theorem_PE_RS}
        The time-averaged reconstruction error under the RS policy is given by:
        \begin{align}
        \label{PE_RS_lemma}
        \scalebox{1}{$
        \!\!P^{\text{RS}}_{E}$} 
            &\!\!=\!
            \scalebox{1}{$
            1\!\!-\!\frac{p_{\alpha_{1}}p_{s_{1/1}}\big(F+2pG+p_{\alpha_{1}}p_{s_{1/1}}\big)}{(p+2q)\!\big[\!\big(2pG-G+1\big)p_{\alpha_{1}}p_{s_{1/1}}\!\!+2q\big(pG-G+1\big)\!\big(1-p_{\alpha_{1}}p_{s_{1/1}}\big)\!\big]}
            $}\notag\\
            &\!-\!
            \scalebox{1}{$
            \frac{2q\big[p+(1-p)p_{\alpha_{1}}p_{\alpha_{2}}p_{s_{2/12}}+(1-p)(1-p_{\alpha_{1}})p_{\alpha_{2}}p_{s_{2/2}}\big]}{(p+2q)\big[3p+(1-3p)p_{\alpha_{1}}p_{\alpha_{2}}p_{s_{2/12}}+(1-3p)(1-p_{\alpha_{1}})p_{\alpha_{2}}p_{s_{2/2}}\big]},
            $}
        \end{align}
        where $F$ and $G$ in \eqref{PE_RS_lemma} are given by:
        \begin{align}
            \label{FG}
            F &\!=\! p_{\alpha_{2}}p_{s_{2/2}}\!\!-\!p_{\alpha_{1}}p_{\alpha_{2}}\!\big[p_{s_{1/1}}\!\!\!-\!p_{s_{1/12}}(1\!\!-\!\!p_{s_{2/2}})\!-\!p_{s_{2/12}}\!\!\!+\!p_{s_{2/2}}\big],\notag\\
            G&\!=\!1-p_{\alpha_{1}}p_{\alpha_{2}}\big[p_{1/12}(1-p_{s_{2/12}})+p_{s_{2/12}}-p_{s_{2/2}}\big]\notag\\
            &-p_{\alpha_{1}}(1-p_{\alpha_{2}})p_{1/1}-p_{\alpha_{2}}p_{s_{2/2}}.
        \end{align}
Moreover, the time-averaged reconstruction error for the error-aware and semantics-aware policies is given by \eqref{PE_EA_lemma}--\eqref{N} and \eqref{PE_SA_lemma}--\eqref{T2}, respectively. Furthermore, the time-averaged reconstruction error for the change-aware policy can be calculated as:
\begin{figure*}[t]
\centering
\begin{align}
\label{PE_EA_lemma}
\!\!\!\!P^{\text{EA}}_{E} &= 
1 \!-\! \frac{1}{Z_{1}} \Big[ q_{\alpha_{1}} p_{s_{1/1}} \big(F' \!\!+\! 2pG'\! \!+\! q_{\alpha_{1}} p_{s_{1/1}} \big) \Big]
\!\!-\!\! \frac{2 p q M q_{\alpha_{1}} p_{s_{1/1}}}{Z_{2}}
\!-\! \frac{2}{Z_{2}} \Big[ q N \!+\! q q_{\alpha_{1}} p_{s_{2/2}} (1\!-\!G') \big( 2q \!+\! (1-2q) q_{\alpha_{1}} p_{s_{1/1}} \big) \Big],\\
\label{F'}
F' &= q_{\alpha_{2}}p_{s_{2/2}} - q_{\alpha_{1}}q_{\alpha_{2}}\big[p_{s_{1/1}} - p_{s_{1/12}}(1-p_{s_{2/2}}) - p_{s_{2/12}} + p_{s_{2/2}}\big],\\
\label{G'}
G' &= 1 - q_{\alpha_{1}} q_{\alpha_{2}} \big[p_{1/12}(1-p_{s_{2/12}}) + p_{s_{2/12}} - p_{s_{2/2}} \big] 
- q_{\alpha_{1}} (1-q_{\alpha_{2}}) p_{1/1} - q_{\alpha_{2}} p_{s_{2/2}},\\
\label{Z1}
Z_1 &= (p + 2q)\Big[\big(2p G' - G' + 1\big) q_{\alpha_{1}} p_{s_{1/1}} + 2q \big(p G' - G' + 1\big)(1 - q_{\alpha_{1}} p_{s_{1/1}})\Big],\\
\label{Z2}
Z_2 &= (p + 2q)\Big[\big(2p G' - G' + 1\big) q_{\alpha_{1}} p_{s_{1/1}} + 2q \big(p G' - G' + 1\big)(1 - q_{\alpha_{1}} p_{s_{1/1}})\Big] 
\big(3p + (1-3p) q_{\alpha_{2}} p_{s_{2/2}}\big),\\
\label{M}
M &= 3 q_{\alpha_{2}} p_{s_{2/2}} (1 - q_{\alpha_{2}} p_{s_{2/2}}) 
+ q_{\alpha_{1}} q_{\alpha_{2}} p_{s_{1/12}} (1 - p_{s_{2/12}}) (1 - 3 q_{\alpha_{2}} p_{s_{2/2}})  + q_{\alpha_{1}} p_{s_{1/1}} (1 - q_{\alpha_{2}}) (1 - 3 q_{\alpha_{2}} p_{s_{2/2}})\notag\\
& 
+ q_{\alpha_{1}} q_{\alpha_{2}} (p_{s_{2/2}} - p_{s_{2/12}}) (-2 + 3 q_{\alpha_{2}} p_{s_{2/2}}),\\
\label{N}
N &= 4 p q q_{\alpha_{2}} p_{s_{2/2}} (1 - q_{\alpha_{1}} p_{s_{1/1}}) (1 - q_{\alpha_{2}} p_{s_{2/2}}) 
+ 2 p^2 G' (1 - q_{\alpha_{2}} p_{s_{2/2}}) \big(q + (1-q) q_{\alpha_{1}} p_{s_{1/1}}\big) \notag\\
& + 2 p q q_{\alpha_{1}} (1 - q_{\alpha_{1}} p_{s_{1/1}}) (1 - 2 q_{\alpha_{2}} p_{s_{2/2}}) 
\big[(1 - q_{\alpha_{2}}) p_{s_{1/1}} - q_{\alpha_{2}} \big(p_{s_{1/12}} (1 - p_{s_{2/12}}) + p_{s_{2/12}} - p_{s_{2/2}}\big)\big].
\end{align}
\hrule height 0.4pt 
\end{figure*}
\begin{figure*}[t]
\centering
\begin{align}
\label{PE_SA_lemma}
P^{\text{SA}}_{E}
&=
\frac{1}{L_{2}}\Big[
6 q p^{3}(1-p_{s_{1/1}})(1-p_{s_{1/12}})(1-p_{s_{2/12}})(1-p_{s_{2/2}})
+ 2 p q p_{s_{2/12}} 
+ 2 p q p_{s_{1/12}} (1-p_{s_{2/12}})\big(4q+(1-4q)p_{s_{2/2}}\big)\notag\\
&
-8 p q p_{s_{1/1}} p_{s_{2/2}} 
+ 2 p q p_{s_{1/1}} p_{s_{2/12}} (1-3p_{s_{2/2}}\!-\!4q\!+\!4q p_{s_{2/2}})
\!+\!  2 p q p_{s_{1/1}} p_{s_{1/12}} (1-p_{s_{2/12}})
\big(2-4q-(3-4q)p_{s_{2/2}}\big)
 \notag\\
&
+8 p q p_{s_{1/1}} \!+\! 6 p q p_{s_{1/12}} 
\!- 14 p q p_{s_{1/1}} p_{s_{2/12}}
- 6 p q p_{s_{1/12}} p_{s_{2/12}}
+ 14 p q p_{s_{1/1}} p_{s_{1/12}} p_{s_{2/12}} 
+ 2 p q p_{s_{2/2}}
- 8 p q p_{s_{1/12}} p_{s_{2/2}}
\notag\\
&
- 8 p q p_{s_{2/12}} p_{s_{2/2}}
+ 16 p q p_{s_{1/1}} p_{s_{2/12}} p_{s_{2/2}}
+ 8 p q p_{s_{1/12}} p_{s_{2/12}} p_{s_{2/2}} 
+ 8 p q^{2}(1-p_{s_{1/1}})(1-p_{s_{1/12}})
(1-p_{s_{2/12}})(1-p_{s_{2/2}})\notag\\
&
+ 6 p q p_{s_{2/12}}- 14 p q p_{s_{1/1}} p_{s_{1/12}} 
+ 16 p q p_{s_{1/1}} p_{s_{1/12}} p_{s_{2/2}} (1-p_{s_{2/12}})
\Big],\\
\label{T2}
        L_{2}&=\Big[2p  (1 - p_{s_{1/12}})  (1 - p_{s_{2/12}})  \big(q + 
      (1 - q)p_{s_{1/1}}  \big)\!+\! \big(p_{s_{1/12}}  (1 - p_{s_{2/12}}) \!+\! p_{s_{2/12}}\big)  \big(2  q + 
      (1 - 2q)p_{s_{1/1}}  \big)\Big]\notag\\
      &\times (p + 2 q)\big(3p-3pp_{s_{2/2}}+p_{s_{2/2}}\big).
\end{align}
\hrule height 0.4pt

\end{figure*}
    \begin{align}
        \label{PE_CA_lemma}
P^{\text{CA}}_{E}&\!\!=1-\frac{2pp_{s_{1/1}}}{Y_{1}}-\frac{2q(1+p_{s_{2/12}})}{(p+2q)(3-p_{s_{2/2}})},
    \end{align}
where $Y_{1}$ in \eqref{PE_CA_lemma} is given by:
     \begin{align}
        \label{Y2}
         Y_{1}\!&=\!\Big[1+p_{s_{1/12}}\!+\!(1-p_{s_{1/12}})\big(p_{s_{2/12}}\!+\!(1-p_{s_{2/12}})p_{s_{1/1}}\big)\Big]\notag\\
         &\times(p+2q).
    \end{align}
	\end{lemma}
   
\begin{IEEEproof} 
See Appendix \ref{Appendix_theorem_PE_RS}.
\end{IEEEproof}
\section{Problem Formulation}
In this section, we formulate and analyze a constrained optimization problem whose objective is to determine the optimal sampling probabilities for the RS and error-aware policies. The aim is to minimize the time-averaged reconstruction error while ensuring that the time-averaged sampling cost remains below a prescribed threshold.

\subsection{Minimizing the time-averaged reconstruction error}
\label{Prob1_PE_RS}
This optimization problem seeks the optimal sampling probabilities $p_{\alpha_{1}}$ and $p_{\alpha_{2}}$ for the RS policy, and $q_{\alpha_{1}}$ and $q_{\alpha_{2}}$ for the error-aware policy, which minimize the time-averaged reconstruction error. We assume that each sampling attempt incurs a cost $\delta$, and that the resulting time-averaged sampling cost must remain below a prescribed threshold $\delta_{\text{max}}$. Focusing first on the RS sampling policy, we formulate the optimization problem as follows:
	\begin{subequations}
		\label{Optimizan_RS}
		\begin{align}
			&\underset{p_{\alpha_{1}},p_{\alpha_{2}}}{\text{minimize}}\hspace{0.1cm}P^{\text{RS}}_{E}\big(p_{\alpha_{1}},p_{\alpha_{2}}\big)\label{Optimization_prob1_objfunc}\\
			&\text{subject to}\!\! \lim_{T \to \infty}\!\frac{1}{T}\mathbb{E}\!\Bigg[\!\sum_{t=1}^{T}\!\delta \!\Big(\!\!\{\alpha_{1}(t)\!=\!1\}\! +\!\mathbbm1\!\{\alpha_{2}(t)\!=\!1\} \!\Big)\! \Bigg]\!\!\leqslant\!\!\delta_{\text{max}}.\label{Optimization_prob1_constraint}
		\end{align}
	\end{subequations}
    \begin{lemma}
		\label{theorem_AvgCost_RS}
The time-averaged sampling cost under the RS policy is given by:
        \begin{align}
        \label{AvgCost_RS_lemma}
            \lim_{T \to \infty}\frac{1}{T}\mathbb{E}\Bigg[\sum_{t=1}^{T}\delta \Big(\mathbbm1\{\alpha_{1}(t)=1\}&\!+\!\mathbbm1\{\alpha_{2}(t)=1\}\Big)\Bigg]\notag\\
            &\!=\!\frac{\delta\big((p+2q)p_{\alpha_{1}}\!+\!2qp_{\alpha_{2}}\big)}{p+2q}.
        \end{align}
	\end{lemma}
    \begin{IEEEproof} 
		See Appendix \ref{Appendix_theorem_timeavg}.
	\end{IEEEproof}
Using Lemma \ref{theorem_AvgCost_RS}, the optimization problem can be simplified as:
\begin{subequations}
		\label{Optimizan_RS2}
		\begin{align}
			&\underset{p_{\alpha_{1}},p_{\alpha_{2}}}{\text{minimize}}\hspace{0.3cm}P^{\text{RS}}_{E}\big(p_{\alpha_{1}},p_{\alpha_{2}}\big)\label{Optimization_prob1_objfunc2}\\
			&\text{subject to}\hspace{0.2cm} \frac{(p+2q)p_{\alpha_{1}}+2qp_{\alpha_{2}}}{p+2q}\leqslant \eta,\label{Optimization_prob1_constraint2}
		\end{align}
	\end{subequations}
 where $\eta=\delta_{\text{max}}/\delta\leqslant1$. Since $p_{s_{1/1}}>p_{s_{1/12}}$ and $p_{s_{2/2}}>p_{s_{2/12}}$, it follows that the objective function in \eqref{Optimization_prob1_objfunc2} is a decreasing function of $p_{\alpha_{2}}$, when $p_{\alpha_{1}}$ is held fixed, i.e., $\frac{\partial P^{\text{RS}}_{E}}{\partial p_{\alpha_2}}<0$. In other words, for any fixed $p_{\alpha_{1}}$, the objective function attains its minimum value when $p_{\alpha_{2}}$ is maximized. Using the constraint in \eqref{Optimization_prob1_constraint2}, the maximum feasible value of $p_{\alpha_{2}}$ is therefore:
    \begin{align}
        \label{maxpa2}
        p_{\alpha_{2}}=\frac{(p+2q)(\eta-p_{\alpha_{1}})}{2q}.
    \end{align}
Using \eqref{maxpa2} and the feasibility condition $0\leqslant p_{\alpha_{2}}\leqslant 1$, the optimization problem can be rewritten as:
   \begin{subequations}
		\label{Optimizan_RS2_2}
		\begin{align}
			&\underset{p_{\alpha_{1}}}{\text{minimize}}\hspace{0.3cm}P^{\text{RS}}_{E}\big(p_{\alpha_{1}}\big)\label{Optimization_prob1_objfunc2_2}\\
			&\text{subject to}\hspace{0.2cm} \max\bigg\{0,\eta-\frac{2q}{p+2q} \bigg\} \leqslant p_{\alpha_{1}} \leqslant \eta.\label{Optimization_prob1_constraint2_2}
		\end{align}
	\end{subequations} 
To determine the value of $p_{\alpha_{1}}$ that minimizes the objective function in \eqref{Optimization_prob1_objfunc2_2}, we first compute the critical points of this function within the feasible interval $\left[\max\left\{0, \eta - \frac{2q}{p + 2q} \right\}, \eta\right]$. The optimal sampling probability $p^{*}_{\alpha_{1}}$ is given by the critical point at which the objective function attains its minimum over this interval. Once $p_{\alpha_{1}}^*$ is obtained, the corresponding optimal sampling probability $p_{\alpha_{2}}^*$ is computed as: 
    \begin{align}
        \label{optimalpa2}
        p^{*}_{\alpha_{2}} = \min \bigg\{1,\frac{(p+2q)(\eta-p^{*}_{\alpha_{1}})}{2q}\bigg\}.
    \end{align}
\begin{remark}
When $p_{\alpha_{1}}=p_{\alpha_{2}}=p_{\alpha}$, the optimization problem in \eqref{Optimizan_RS2} reduces to the following single-variable formulation:
\begin{subequations}
		\label{Optimizan_RS_same_pa}
		\begin{align}
			&\underset{p_{\alpha}}{\text{minimize}}\hspace{0.3cm}P^{\text{RS}}_{E}\big(p_{\alpha}\big)\label{Optimization_prob1_objfunc_pa}\\
			&\text{subject to}\hspace{0.2cm} p_{\alpha}\leqslant \frac{(p+2q)\eta}{p+4q}.\label{Optimization_prob1_constraint_pa}
		\end{align}
	\end{subequations}
Since the objective function in \eqref{Optimization_prob1_objfunc_pa} is strictly decreasing in $p_{\alpha}$, its minimum is achieved by choosing the largest feasible sampling probability. From the constraint in \eqref{Optimization_prob1_constraint_pa}, the maximum allowable value of $p_{\alpha}$ is $\frac{(p+2q)\eta}{p+4q}$. Therefore, the optimal sampling probability is $p^{*}_{\alpha} = \frac{(p+2q)\eta}{p+4q}$.
\end{remark}
Now, using \eqref{PE_EA_lemma}, the corresponding optimization problem for the error-aware policy can be formulated as follows:
	\begin{subequations}
		\label{Optimizan_EA}
		\begin{align}
			&\underset{q_{\alpha_{1}},q_{\alpha_{2}}}{\text{minimize}}\hspace{0.1cm}P^{\text{EA}}_{E}\big(q_{\alpha_{1}},q_{\alpha_{2}}\big)\label{Optimization_prob1_objfunc_EA}\\
			&\text{subject to}\!\! \lim_{T \to \infty}\!\frac{1}{T}\mathbb{E}\!\Bigg[\!\sum_{t=1}^{T}\!\delta \!\Big(\!\mathbbm1\!\{\alpha_{1}(t)\!=\!1\}\! +\!\mathbbm1\!\{\alpha_{2}(t)\!=\!1\} \!\Big)\! \Bigg]\!\!\leqslant\!\!\delta_{\text{max}}.\label{Optimization_prob1_constraint_EA}
		\end{align}
	\end{subequations}
\begin{lemma}
\label{theorem_AvgCost_EA}
The time-averaged sampling cost under the error-aware policy is given by:
\begin{align}
      \label{TAvg_EA_lemma}
     &\!\scalebox{0.9}{$\displaystyle{\lim_{T \to \infty}\frac{1}{T}}\mathbb{E}\Bigg[\displaystyle{\sum_{t=1}^{T}}\delta\Big(\mathbbm1\{\alpha_{1}(t)=1\}+\mathbbm1\{\alpha_{2}(t)=1\}\Big) \Bigg]$} \notag\\    
     &\!\!=\scalebox{0.9}{$\displaystyle{\frac{2pq\delta}{Z_{2}}}\Big[3  G'  p^2  q_{\alpha_{1}}  (1 - q_{\alpha_{2}}  p_{s_{2/2}})+q_{\alpha_{1}}  q^{2}_{\alpha_{2}}  p_{s_{2/2}}  (2 p_{s_{1/1}} + p_{s_{2/2}})$}\notag\\
     &\!\!+\!\scalebox{0.9}{$2q^{2}_{\alpha_{1}}q_{\alpha_{2}}  (1 \!\!-\!\! q_{\alpha_{2}})  p_{s_{1/1}}^2\!\!\!+\!\!(q_{\alpha_{1}}q_{\alpha_{2}})^{2}  p_{s_{1/1}}\!  \big(2  p_{s_{1/12}}  (1 \!-\! p_{s_{2/12}}) \!+\! p_{s_{2/12}}\big)$}\notag\\
     &\!\!+\scalebox{0.9}{$(q_{\alpha_{1}}q_{\alpha_{2}})^{2}  p_{s_{2/2}}  \big(p_{s_{1/12}}  (1 - p_{s_{2/12}}) + p_{s_{2/12}} - p_{s_{2/2}}\big)$}\notag\\
     &\!\!+\scalebox{0.9}{$2q^{2}_{\alpha_{1}}q_{\alpha_{2}}  (1 - q_{\alpha_{2}})  p_{s_{1/1}}  p_{s_{2/2}}+4qq_{\alpha_{2}}  (1 - G')    (1 - q_{\alpha_{1}}  p_{s_{1/1}})$}\notag\\
     &\!\!+\scalebox{0.9}{$4pq  q_{\alpha_{2}}  (1 - q_{\alpha_{2}}  p_{s_{2/2}}) +4  p  q_{\alpha_{1}}  q_{\alpha_{2}} (p_{s_{1/1}} + p_{s_{2/2}})  (1 - q_{\alpha_{2}}  p_{s_{2/2}})$}\notag\\
     &\!\!-\scalebox{0.9}{$4p  q_{\alpha_{1}}  q^{2}_{\alpha_{2}}  \big(p_{s_{1/12}}  (1 - p_{s_{2/12}}) + p_{s_{2/12}} - p_{s_{2/2}}\big)-8  p  q  q_{\alpha_{1}}  q_{\alpha_{2}}  p_{s_{1/1}}$}\notag\\
     &\!\!-\scalebox{0.9}{$p  q_{\alpha_{1}}^2 q_{\alpha_{2}}  (p_{s_{1/12}}  (1 - p_{s_{2/12}}) + p_{s_{2/12}} - p_{s_{2/2}})  (-3 + 4  q_{\alpha_{2}}  p_{s_{2/2}})$}\notag\\
     &\!\!+\scalebox{0.9}{$4  p  q  q_{\alpha_{1}}  q^{2}_{\alpha_{2}}   p_{s_{1/1}}  (1 +  p_{s_{2/2}}) -4  p (1 - q)  q_{\alpha_{1}}^2  q_{\alpha_{2}}(1 - q_{\alpha_{2}})   p_{s_{1/1}}^2$} \notag\\
     &\!\!-\!\scalebox{0.9}{$3pq_{\alpha_{1}}^{2}q_{\alpha_{2}}p_{s_{1/1}}\!\!\!-\!\!7pq_{\alpha_{1}}^{2}q_{\alpha_{2}}p_{s_{1/1}}p_{s_{2/2}}\!\!-\!\!p(q_{\alpha_{1}}q_{\alpha_{2}})^{2}p_{s_{1/1}}p_{s_{2/12}}$}\notag\\
     &\!\!-\!\scalebox{0.9}{$p(q_{\alpha_{1}}q_{\alpha_{2}})^{2}p_{s_{1/1}}\big(-5  p_{s_{2/2}} \!+\! 4  p_{s_{1/12}}  (1 \!-\! p_{s_{2/12}})  (1 \!-\! q) \!-\! 4q p_{s_{2/12}} \big)$}\notag\\
     &\!\!+\scalebox{0.9}{$6pq_{\alpha_{1}}^{2}p_{s_{1/1}}-4pq(q_{\alpha_{1}}q_{\alpha_{2}})^{2}p_{s_{1/1}}p_{s_{2/2}}\Big]$},
 \end{align}
 where $G'$ and $Z_{2}$  were given in \eqref{G'} and \eqref{Z2}.
	\end{lemma}
    \begin{IEEEproof} 
		See Appendix \ref{Appendix_theorem_timeavg}.
	\end{IEEEproof}

This optimization problem is non-convex in the variables $q_{\alpha_{1}}$ and $q_{\alpha_{2}}$, and, as indicated by \eqref{TAvg_EA_lemma}, the optimization parameters are interdependent. Consequently, obtaining a closed-form solution is cumbersome, if not infeasible, and we therefore solve \eqref{Optimizan_EA} numerically.
\section{Numerical Results}
In this section, we numerically evaluate our analytical results and examine the performance of the proposed sampling policies in terms of the time-averaged reconstruction error under various system parameter settings.
\begin{figure*}[!t]
    \centering
    \footnotesize

\subfigure[$p_{s_{1/1}}\!=\!0.2, p_{s_{1/12}}\!=\!0.1,$
$p_{s_{2/2}}\!=\!0.2, p_{s_{2/12}}\!=\!0.1.$]{%
\includegraphics[trim=0.5cm 0.05cm 1.1cm 0.6cm,
width=0.45\textwidth, clip]{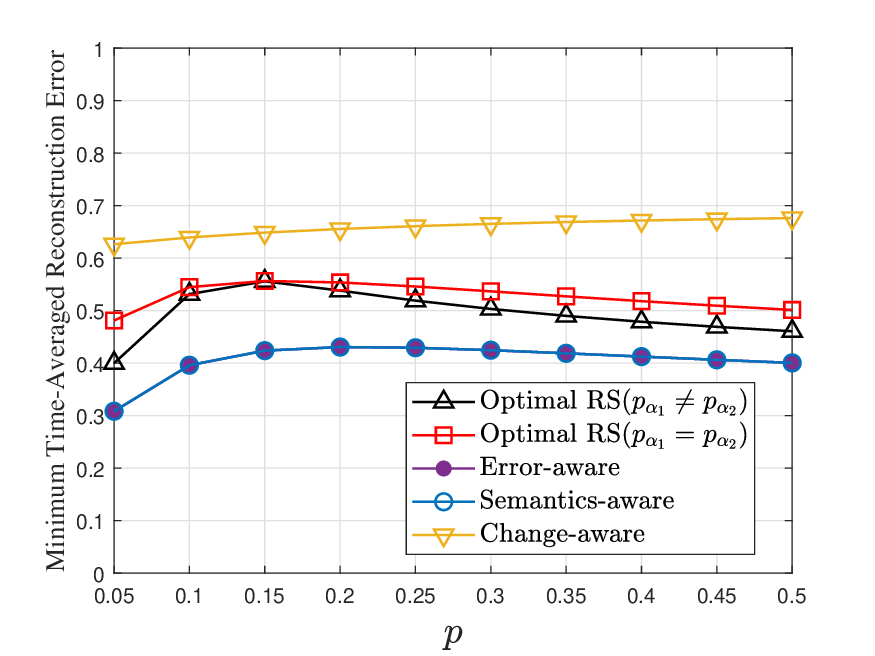}}
\hfill
\subfigure[$p_{s_{1/1}}\!=\!0.8, p_{s_{1/12}}\!=\!0.1,$
$p_{s_{2/2}}\!=\!0.2, p_{s_{2/12}}\!=\!0.1.$]{%
\includegraphics[trim=0.5cm 0.05cm 1.1cm 0.6cm,
width=0.45\textwidth, clip]{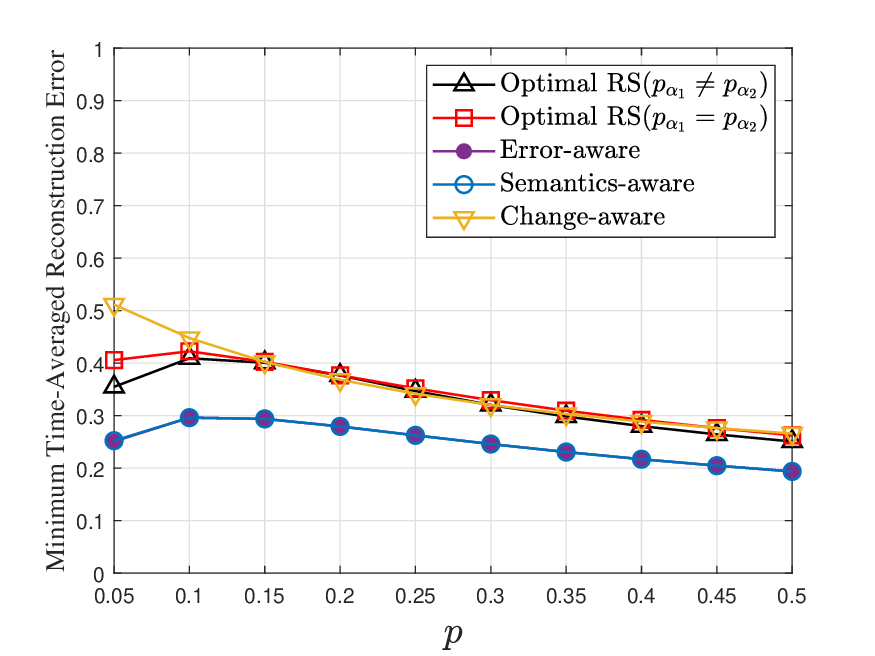}} \\[0.35cm]

\subfigure[$p_{s_{1/1}}\!=\!0.2, p_{s_{1/12}}\!=\!0.1,$ 
$p_{s_{2/2}}\!=\!0.8, p_{s_{2/12}}\!=\!0.1.$]{%
\includegraphics[trim=0.5cm 0.05cm 1.1cm 0.6cm,
width=0.45\textwidth, clip]{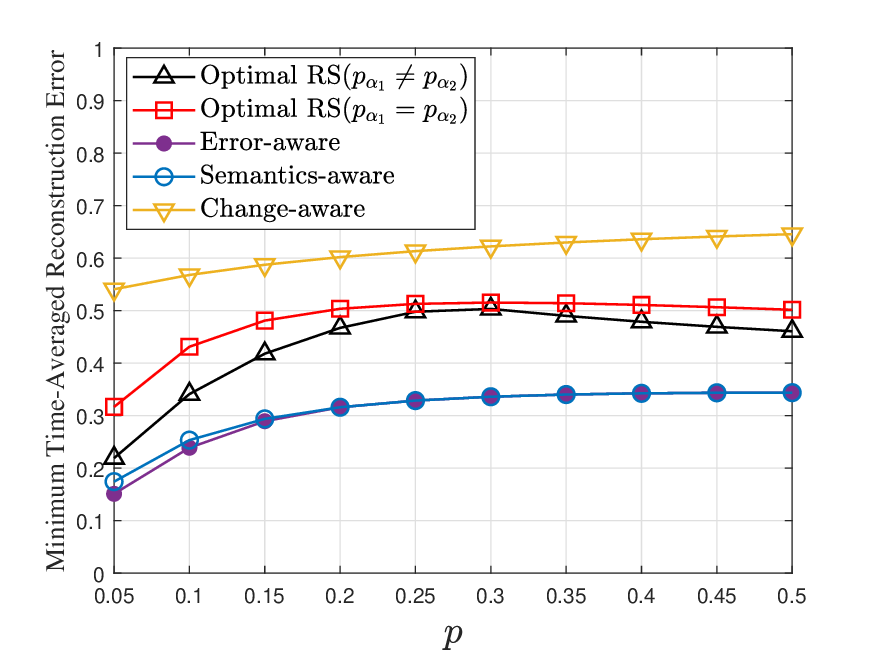}}
\hfill
\subfigure[$p_{s_{1/1}}\!=\!0.8, p_{s_{1/12}}\!=\!0.1,$ 
$p_{s_{2/2}}\!=\!0.8, p_{s_{2/12}}\!=\!0.1.$]{%
\includegraphics[trim=0.5cm 0.05cm 1.1cm 0.6cm,
width=0.45\textwidth, clip]{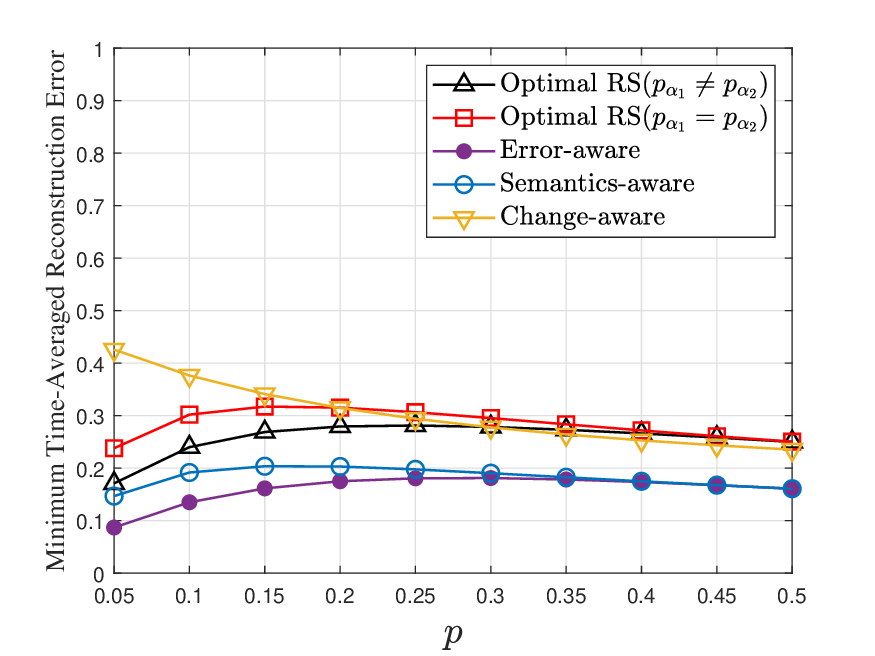}}

\caption{Minimum time-averaged reconstruction error as a function of $p$ for $\eta=0.8$, $q=0.1$ and the success probabilities. For system parameters that violate the cost constraint, the change-aware and semantics-aware policies do not admit feasible solutions; therefore, these cases are excluded from the figures.}
\label{TAvgReconstructionerror_PQVAr1}
\end{figure*}
\begin{figure*}[!t]
    \centering
    \footnotesize

\subfigure[$p_{s_{1/1}}\!=\!0.2, p_{s_{1/12}}\!=\!0.1,$
$p_{s_{2/2}}\!=\!0.2, p_{s_{2/12}}\!=\!0.1.$]{%
\includegraphics[trim=0.5cm 0.05cm 1.1cm 0.6cm,
width=0.45\textwidth, clip]{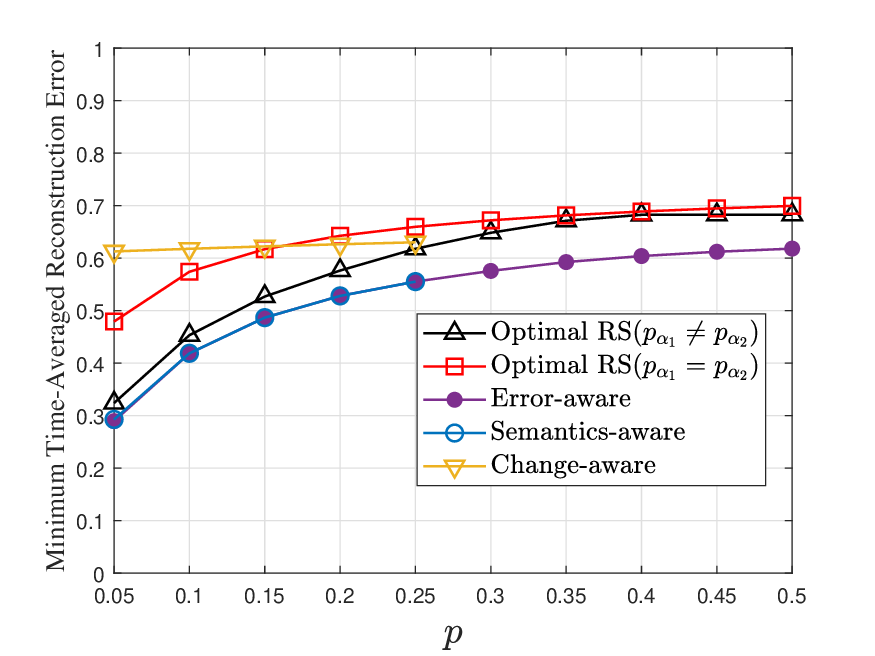}}
\hfill
\subfigure[$p_{s_{1/1}}\!=\!0.8, p_{s_{1/12}}\!=\!0.1,$ 
$p_{s_{2/2}}\!=\!0.2, p_{s_{2/12}}\!=\!0.1.$]{%
\includegraphics[trim=0.5cm 0.05cm 1.1cm 0.6cm,
width=0.45\textwidth, clip]{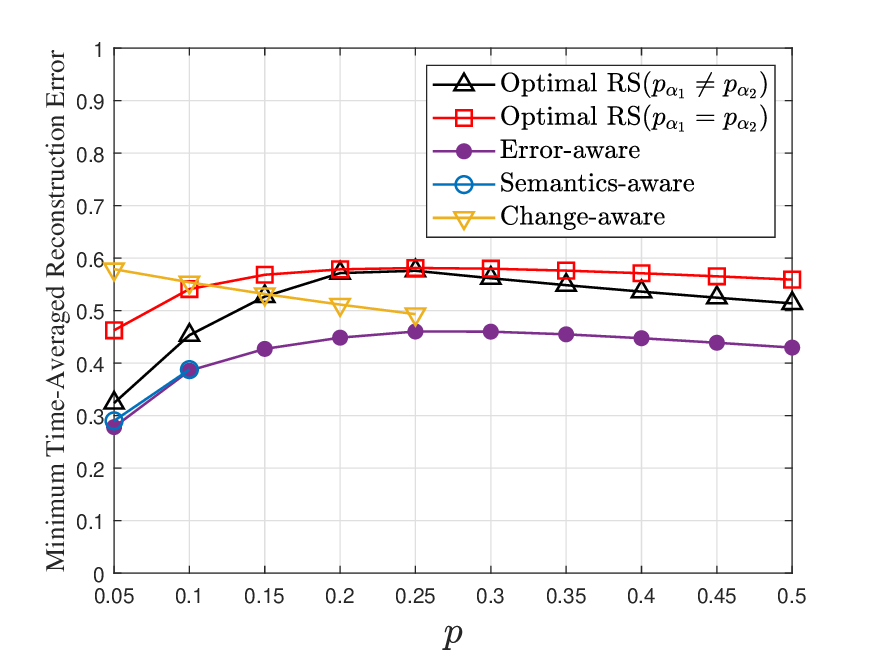}} \\[0.35cm]

\subfigure[$p_{s_{1/1}}\!=\!0.2, p_{s_{1/12}}\!=\!0.1,$ 
$p_{s_{2/2}}\!=\!0.8, p_{s_{2/12}}\!=\!0.1.$]{%
\includegraphics[trim=0.5cm 0.05cm 1.1cm 0.6cm,
width=0.45\textwidth, clip]{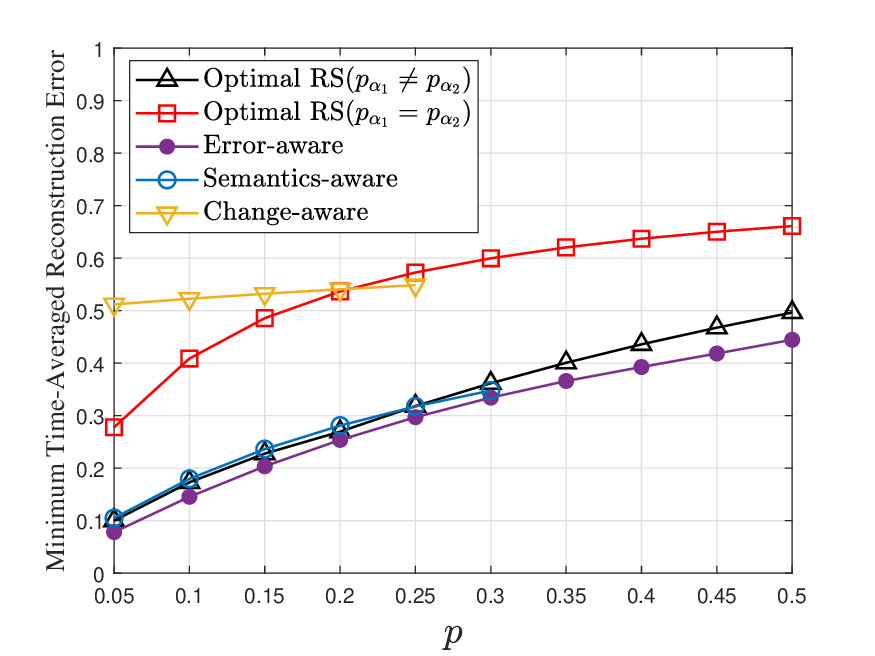}}
\hfill
\subfigure[ $p_{s_{1/1}}\!=\!0.8, p_{s_{1/12}}\!=\!0.1,$
$p_{s_{2/2}}\!=\!0.8, p_{s_{2/12}}\!=\!0.1.$]{%
\includegraphics[trim=0.5cm 0.05cm 1.1cm 0.6cm,
width=0.45\textwidth, clip]{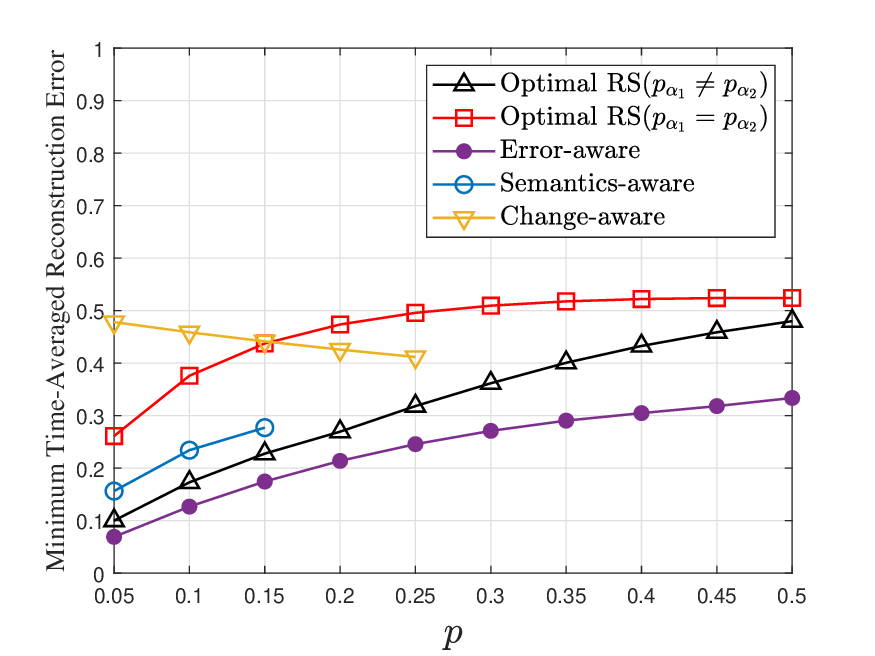}}

\caption{Minimum time-averaged reconstruction error as a function of $p$ for $\eta=0.8$, $q=0.1$ and the success probabilities. For system parameters that violate the cost constraint, the change-aware and semantics-aware policies do not admit feasible solutions; therefore, these cases are excluded from the figures.}
\label{TAvgReconstructionerror_PQVAr2}
\end{figure*}
\par
\begin{figure*}[!t]
    \centering
    \footnotesize

\subfigure[$p_{s_{1/1}}\!=\!0.2, p_{s_{1/12}}\!=\!0.1,$ 
$p_{s_{2/2}}\!=\!0.2, p_{s_{2/12}}\!=\!0.1.$]{%
\includegraphics[trim=0.5cm 0.05cm 1.1cm 0.6cm,
width=0.45\textwidth, clip]{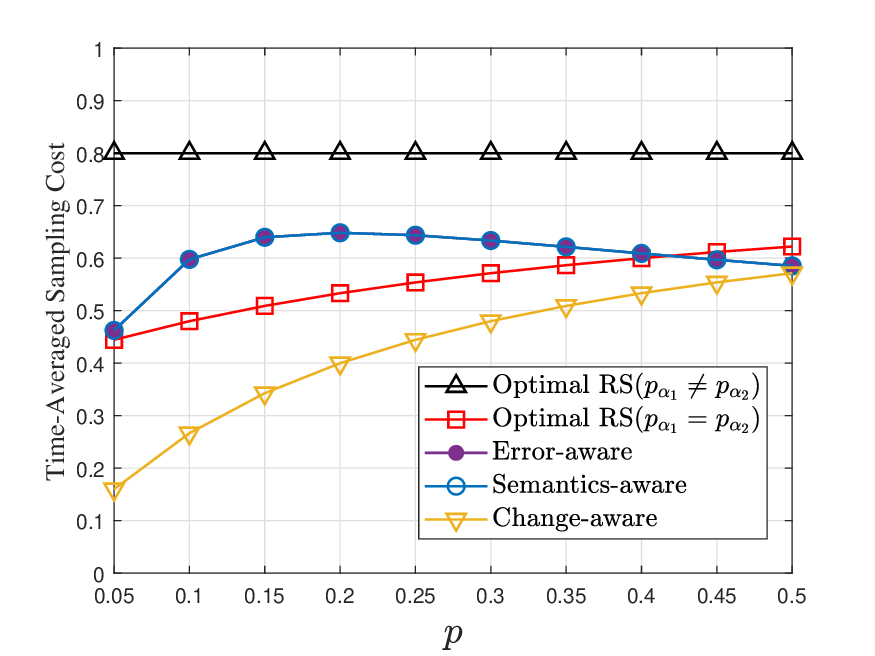}
\label{TAvgSamplingCost1_q0.1}}
\hfill
\subfigure[$p_{s_{1/1}}\!=\!0.8, p_{s_{1/12}}\!=\!0.1,$
$p_{s_{2/2}}\!=\!0.2, p_{s_{2/12}}\!=\!0.1.$]{%
\includegraphics[trim=0.5cm 0.05cm 1.1cm 0.6cm,
width=0.45\textwidth, clip]{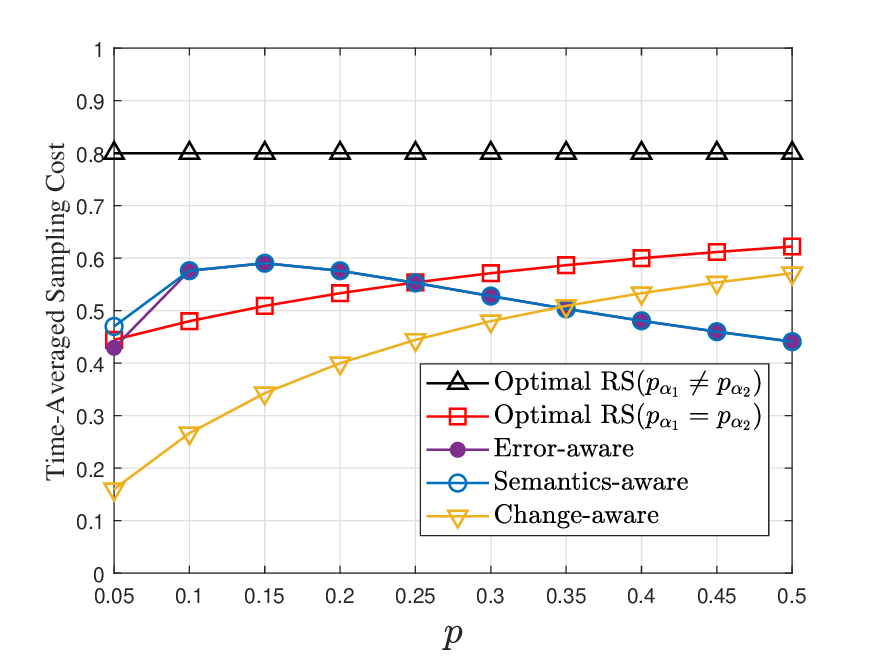}
\label{TAvgSamplingCost2_q0.1}} \\[0.35cm]

\subfigure[$p_{s_{1/1}}\!=\!0.2, p_{s_{1/12}}\!=\!0.1,$
$p_{s_{2/2}}\!=\!0.8, p_{s_{2/12}}\!=\!0.1.$]{%
\includegraphics[trim=0.5cm 0.05cm 1.1cm 0.6cm,
width=0.45\textwidth, clip]{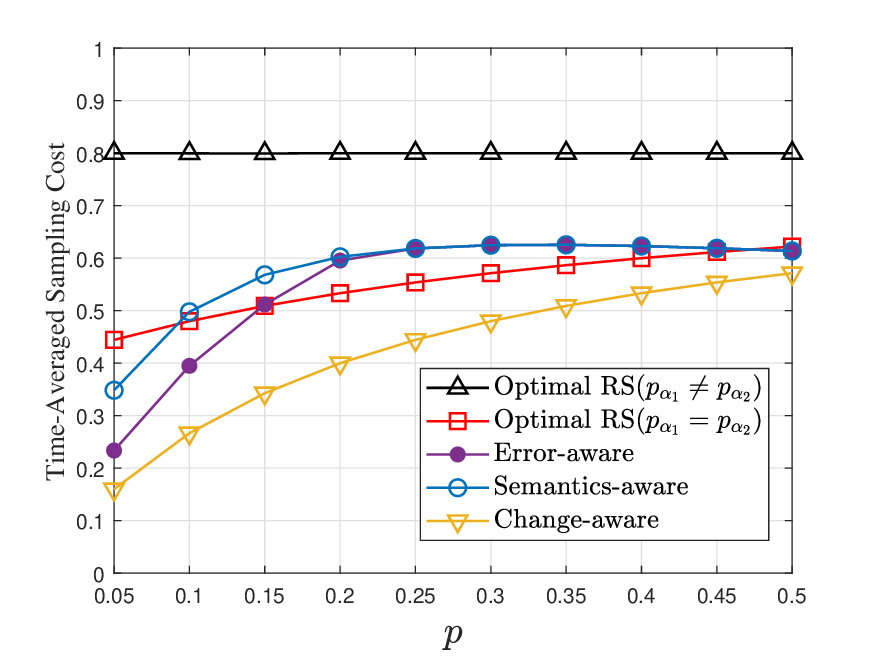}
\label{TAvgSamplingCost3_q0.1}}
\hfill
\subfigure[$p_{s_{1/1}}\!=\!0.8, p_{s_{1/12}}\!=\!0.1,$ 
$p_{s_{2/2}}\!=\!0.8, p_{s_{2/12}}\!=\!0.1.$]{%
\includegraphics[trim=0.5cm 0.05cm 1.1cm 0.6cm,
width=0.45\textwidth, clip]{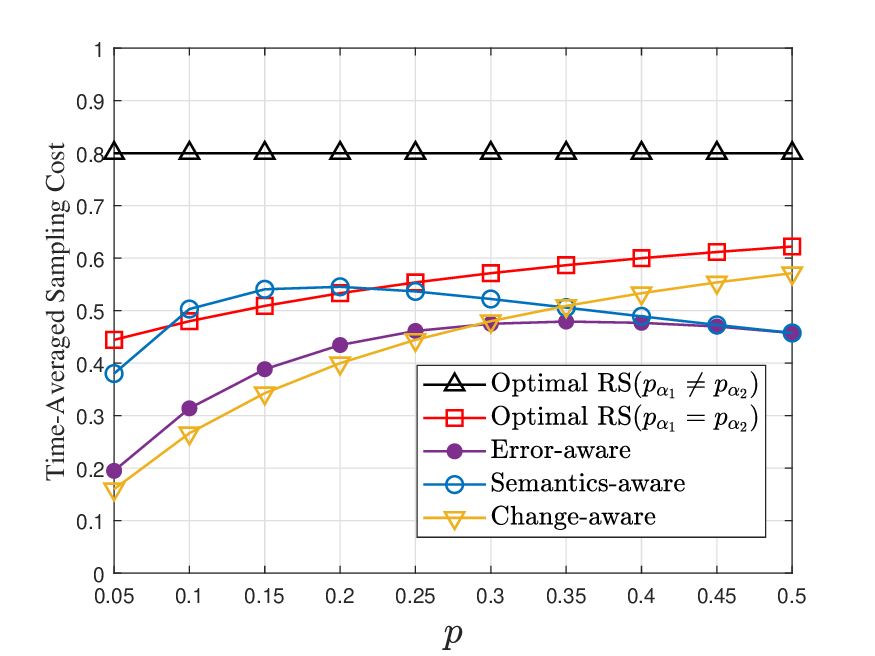}
\label{TAvgSamplingCost4_q0.1}}

\caption{Time-averaged sampling cost as a function of $p$, for $\eta=0.8$, $q=0.1$ and the success probabilities.}
\label{TAvgSamplingCost_PQVAR1}
\end{figure*}

\begin{figure*}[!t]
    \centering
    \footnotesize

\subfigure[$p_{s_{1/1}}\!=\!0.2, p_{s_{1/12}}\!=\!0.1,$ 
$p_{s_{2/2}}\!=\!0.2, p_{s_{2/12}}\!=\!0.1.$]{%
\includegraphics[trim=0.5cm 0.05cm 1.1cm 0.6cm,
width=0.45\textwidth, clip]{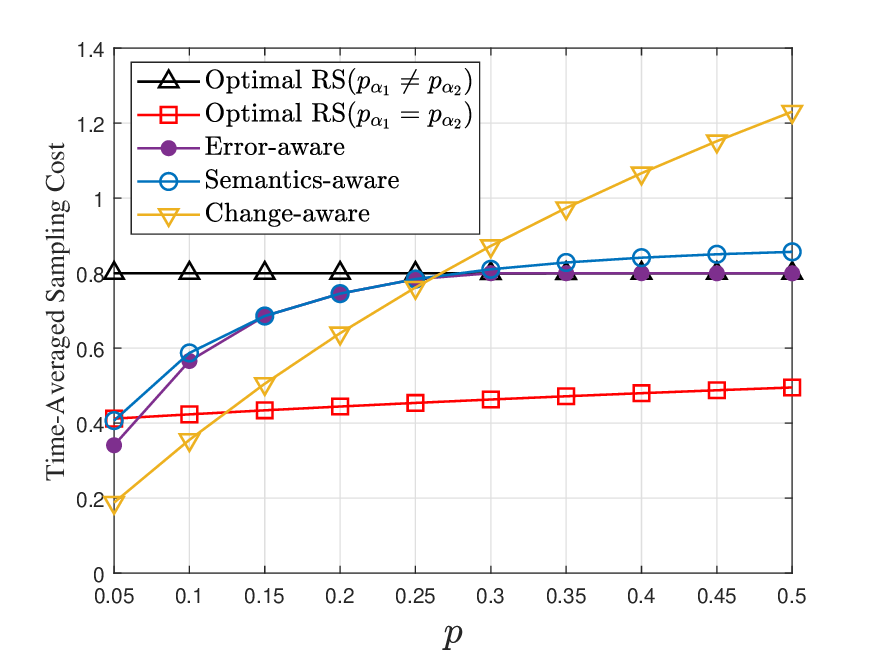}
\label{TAvgSamplingCost1_q0.4}}
\hfill
\subfigure[$p_{s_{1/1}}\!=\!0.4, p_{s_{1/12}}\!=\!0.1,$ 
$p_{s_{2/2}}\!=\!0.2, p_{s_{2/12}}\!=\!0.1.$]{%
.\includegraphics[trim=0.5cm 0.05cm 1.1cm 0.6cm,
width=0.45\textwidth, clip]{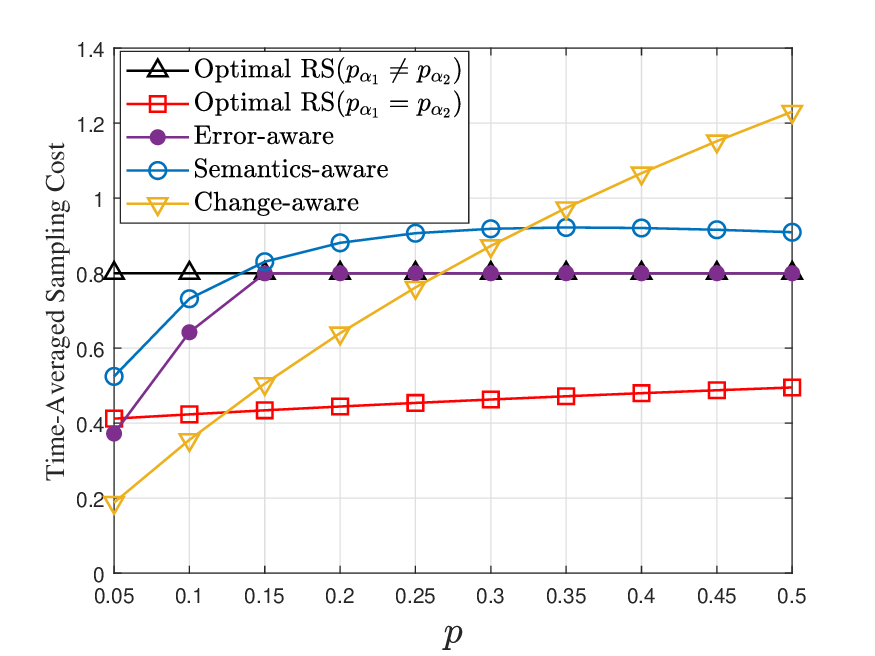}
\label{TAvgSamplingCost2_q0.4}} \\[0.35cm]

\subfigure[$p_{s_{1/1}}\!=\!0.2, p_{s_{1/12}}\!=\!0.1,$
$p_{s_{2/2}}\!=\!0.8, p_{s_{2/12}}\!=\!0.1.$]{%
\includegraphics[trim=0.5cm 0.05cm 1.1cm 0.6cm,
width=0.45\textwidth, clip]{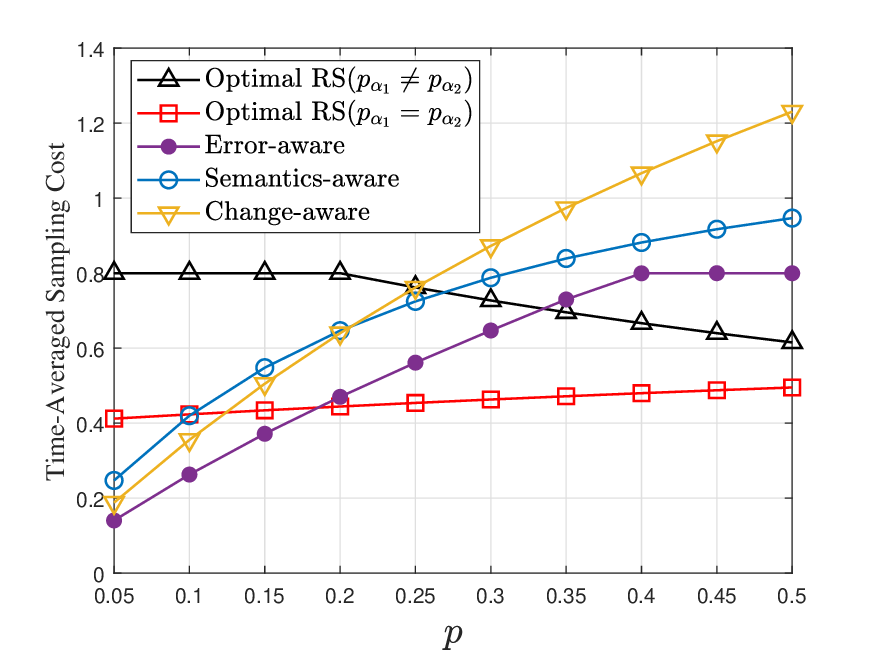}
\label{TAvgSamplingCost3_q0.4}}
\hfill
\subfigure[$p_{s_{1/1}}\!=\!0.8, p_{s_{1/12}}\!=\!0.1,$ 
$p_{s_{2/2}}\!=\!0.8, p_{s_{2/12}}\!=\!0.1.$]{%
\includegraphics[trim=0.5cm 0.05cm 1.1cm 0.6cm,
width=0.45\textwidth, clip]{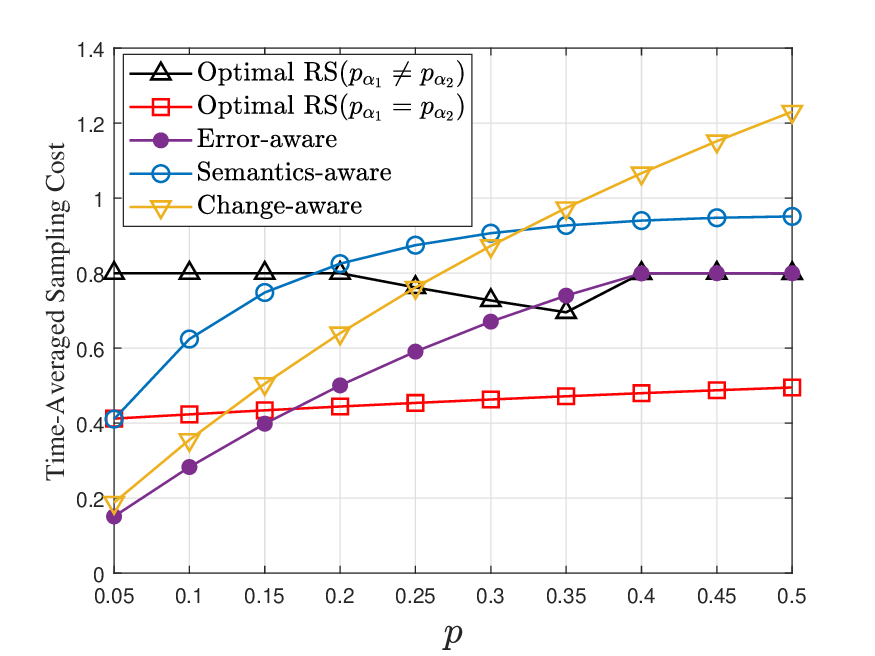}
\label{TAvgSamplingCost4_q0.4}}

\caption{Time-averaged sampling cost as a function of $p$, for $\eta=0.8$, $q=0.4$ and the success probabilities.}
\label{TAvgSamplingCost_PQVAR2}
\end{figure*}

 Figs.~\ref{TAvgReconstructionerror_PQVAr1} and \ref{TAvgReconstructionerror_PQVAr2} illustrate the minimum time-averaged reconstruction error under a sampling cost constraint as a function of $p$, for $\eta = 0.8$, $q=0.4$ and success probabilities. \emph{The results show that the optimal error-aware policy consistently outperforms the other policies across different levels of statistical correlation between $X_1$ and $X_2$, regardless of whether the success probabilities are low or high}. For each fixed value of $q$, increasing $p$ raises the marginal probability $\mathbb{P}[X_1 = 0]$ while decreasing the joint probabilities $\mathbb{P}[X_1 = 1, X_2 = 0]$ and $\mathbb{P}[X_1 = 1, X_2 = 1]$. In this regime, $X_2$ becomes more conditionally dependent on $X_1$, allowing the receiver to decode $X_2$ more reliably using information from $X_1$. Consequently, the sampling probabilities $p_{\alpha_1}$ (for the RS policy) and $q_{\alpha_1}$ (for the error-aware policy) increase. Conversely, when $q$ rises, the joint probabilities $\mathbb{P}[X_1 = 1, X_2 = 0]$ and $\mathbb{P}[X_1 = 1, X_2 = 1]$ increase, making $X_1$ more conditionally dependent on $X_2$. In this case, the sampling probabilities $p_{\alpha_2}$ and $q_{\alpha_2}$ also increase relative to scenarios with smaller $q$. 
\emph{These results confirm that the statistical correlation between $X_1$ and $X_2$ directly shapes the optimal sampling decisions in both policies}. The error-aware policy triggers sampling probabilistically only when at least one receiver is in an erroneous state, requiring fewer sampling actions than the change-aware and semantics-aware policies. However, the latter two may exceed the sampling cost constraint, particularly when $p$ and $q$ are high, conditions that indicate \emph{strong correlation} between the two processes. \emph{In such cases, since information about each process can be inferred from the other with reduced uncertainty, achieving the minimum time-averaged reconstruction error requires more frequent sampling, which may violate the time-averaged cost budget}. This property highlights the error-aware policy's suitability for resource-constrained environments such as energy-limited wireless networks. Moreover, the optimal error-aware policy avoids unnecessary sampling and achieves a higher sampling probability than the optimal RS policy, while maintaining a similar or lower time-averaged sampling cost, as shown in Figs.~\ref{TAvgSamplingCost_PQVAR1} and \ref{TAvgSamplingCost_PQVAR2}. Additionally, for certain values of $q$, $p$, $p_{s_{1/1}}$, $p_{s_{1/12}}$, $p_{s_{2/2}}$, and $p_{s_{2/12}}$ that violate the cost constraint, no feasible solution exists for the change-aware and semantics-aware policies, and these cases are therefore excluded from Figs. ~\ref{TAvgReconstructionerror_PQVAr1} and \ref{TAvgReconstructionerror_PQVAr2}. In contrast, the RS and error-aware policies are capable of adjusting the sampling probabilities to satisfy the constraint and minimize the time-averaged reconstruction error. 

\begin{figure*}[!t]
    \centering
    \footnotesize

\subfigure[$p_{s_{1/1}}\!=\!0.2,$
$p_{s_{1/12}}\!=\!0.1, p_{s_{2/2}}\!=\!0.2, p_{s_{2/12}}\!=\!0.1.$]{%
\includegraphics[trim=0.5cm 0.05cm 1.1cm 0.6cm,
width=0.45\textwidth, clip]{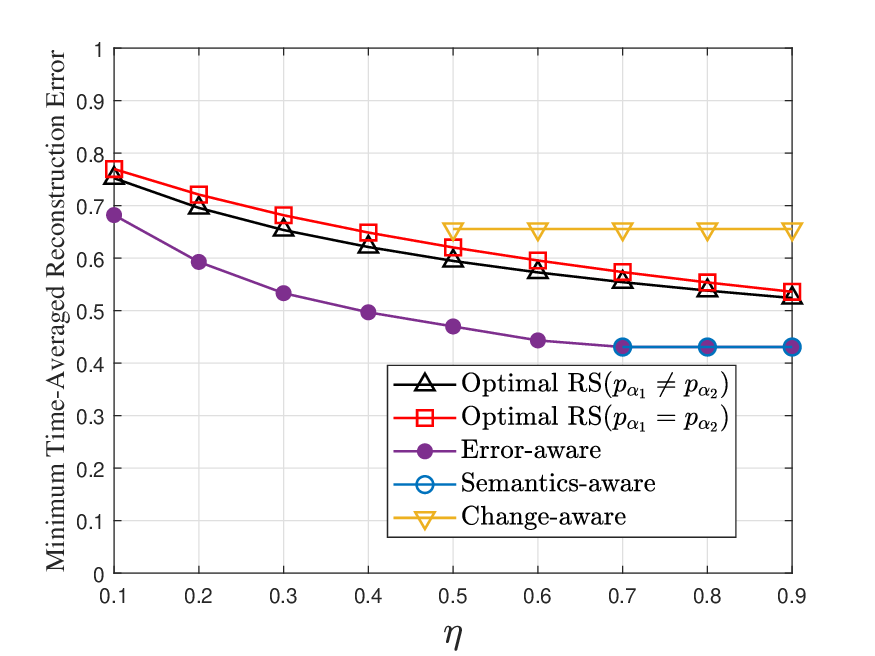}}
\hfill
\subfigure[$p_{s_{1/1}}\!=\!0.8,$
$p_{s_{1/12}}\!=\!0.1, p_{s_{2/2}}\!=\!0.2, p_{s_{2/12}}\!=\!0.1.$]{%
\includegraphics[trim=0.5cm 0.05cm 1.1cm 0.6cm,
width=0.45\textwidth, clip]{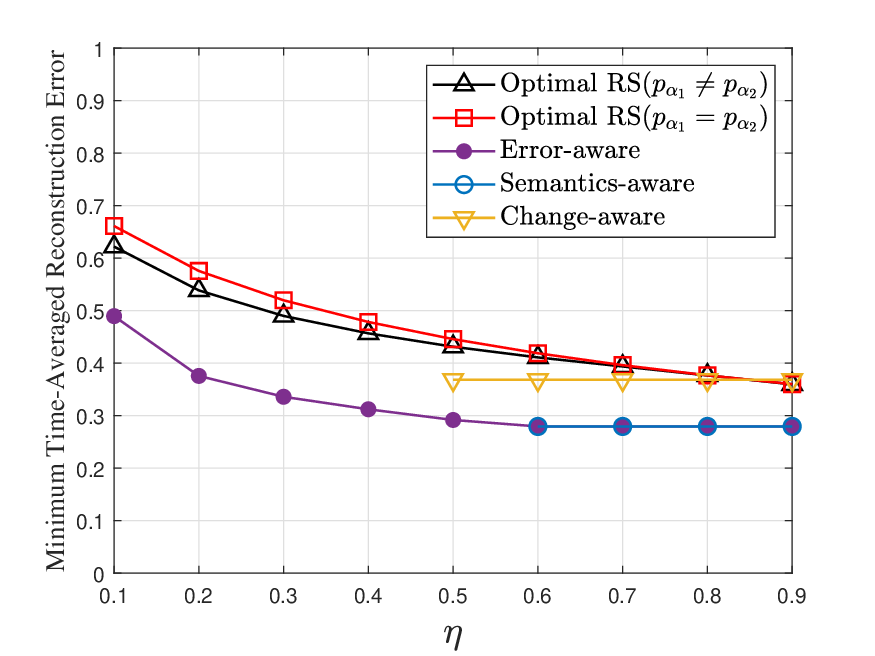}} \\[0.35cm]

\subfigure[$p_{s_{1/1}}\!=\!0.2,$ 
$p_{s_{1/12}}\!=\!0.1, p_{s_{2/2}}\!=\!0.8, p_{s_{2/12}}\!=\!0.1.$]{%
\includegraphics[trim=0.5cm 0.05cm 1.1cm 0.6cm,
width=0.45\textwidth, clip]{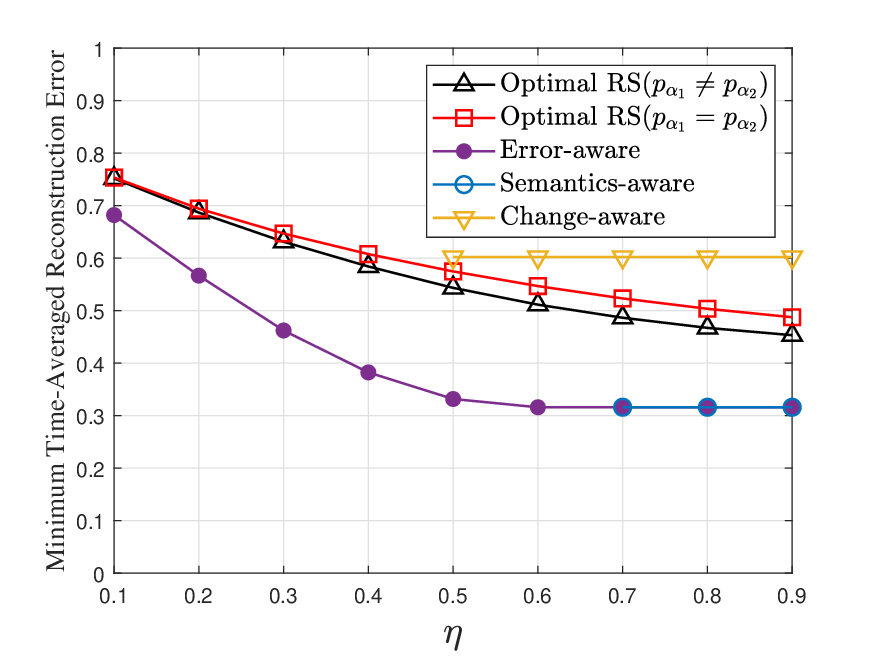}}
\hfill
\subfigure[$ p_{s_{1/1}}\!=\!0.8,$ 
$p_{s_{1/12}}\!=\!0.1, p_{s_{2/2}}\!=\!0.8, p_{s_{2/12}}\!=\!0.1.$]{%
\includegraphics[trim=0.5cm 0.05cm 1.1cm 0.6cm,
width=0.45\textwidth, clip]{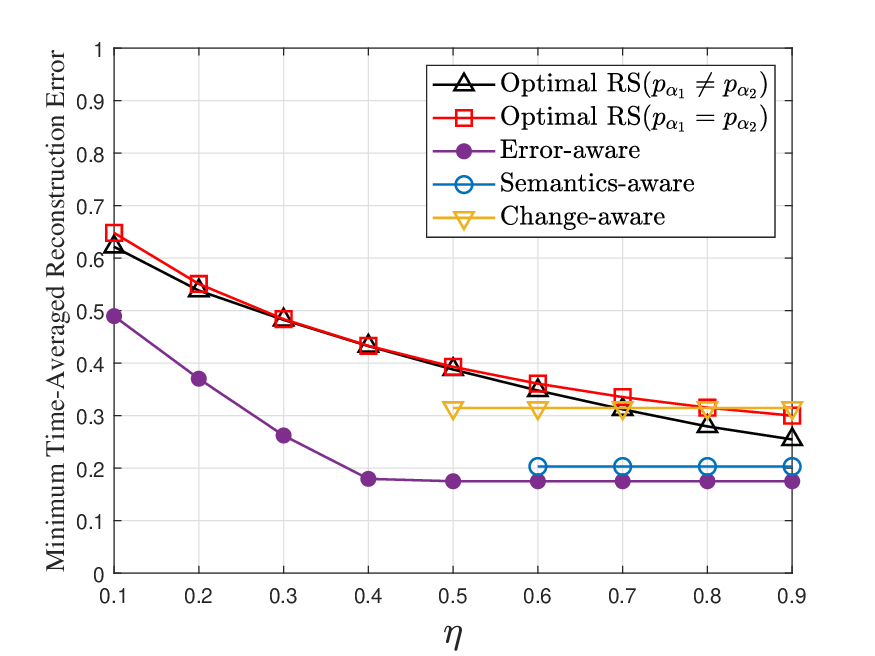}}

\caption{Minimum time-averaged reconstruction error as a function of $\eta$ for $p=0.2$, $q=0.1$, and success probabilities. For system parameters that violate the cost constraint, the change-aware and semantics-aware policies do not admit feasible solutions; therefore, these cases are excluded from the figures.}
\label{TAvgReconstructionerror_PQFix_ETAVAr1}
\end{figure*}

\begin{figure*}[!t]
    \centering
    \footnotesize

\subfigure[$p_{s_{1/1}}\!=\!0.2,$ 
$p_{s_{1/12}}\!=\!0.1, p_{s_{2/2}}\!=\!0.2, p_{s_{2/12}}\!=\!0.1.$]{%
\includegraphics[trim=0.5cm 0.05cm 1.1cm 0.6cm,
width=0.45\textwidth, clip]{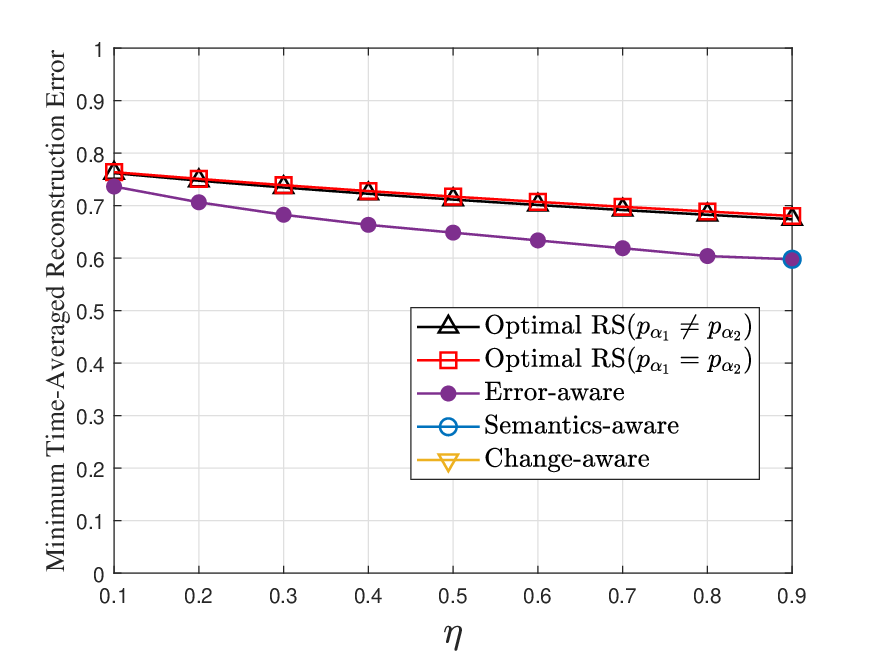}}
\hfill
\subfigure[$p_{s_{1/1}}\!=\!0.8,$ 
$p_{s_{1/12}}\!=\!0.1, p_{s_{2/2}}\!=\!0.2, p_{s_{2/12}}\!=\!0.1.$]{%
\includegraphics[trim=0.5cm 0.05cm 1.1cm 0.6cm,
width=0.45\textwidth, clip]{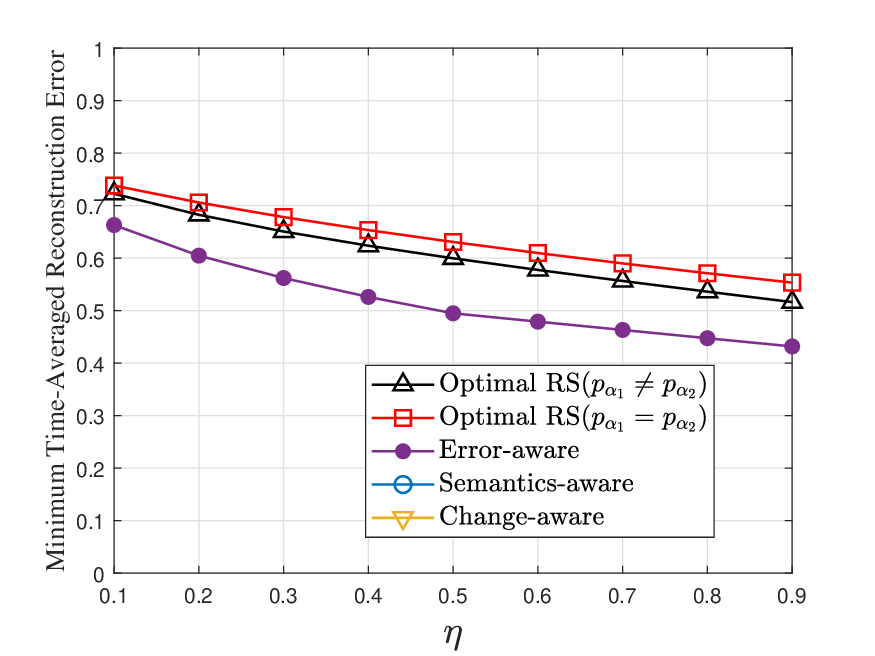}} \\[0.35cm]

\subfigure[$p_{s_{1/1}}\!=\!0.2,$ 
$p_{s_{1/12}}\!=\!0.1, p_{s_{2/2}}\!=\!0.8, p_{s_{2/12}}\!=\!0.1.$]{%
\includegraphics[trim=0.5cm 0.05cm 1.1cm 0.6cm,
width=0.45\textwidth, clip]{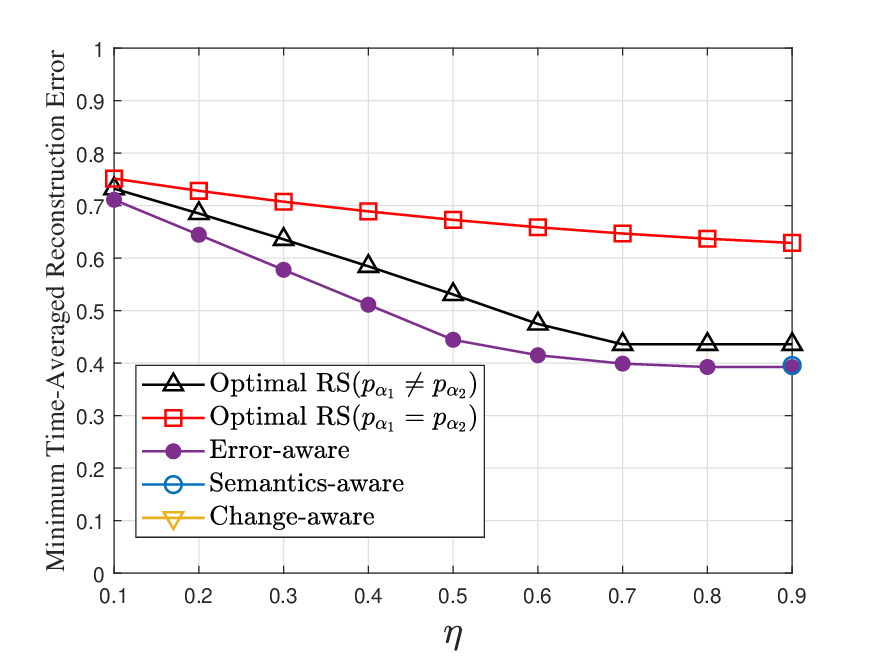}}
\hfill
\subfigure[$p_{s_{1/1}}\!=\!0.8,$
$p_{s_{1/12}}\!=\!0.1, p_{s_{2/2}}\!=\!0.8, p_{s_{2/12}}\!=\!0.1.$]{%
\includegraphics[trim=0.5cm 0.05cm 1.1cm 0.6cm,
width=0.45\textwidth, clip]{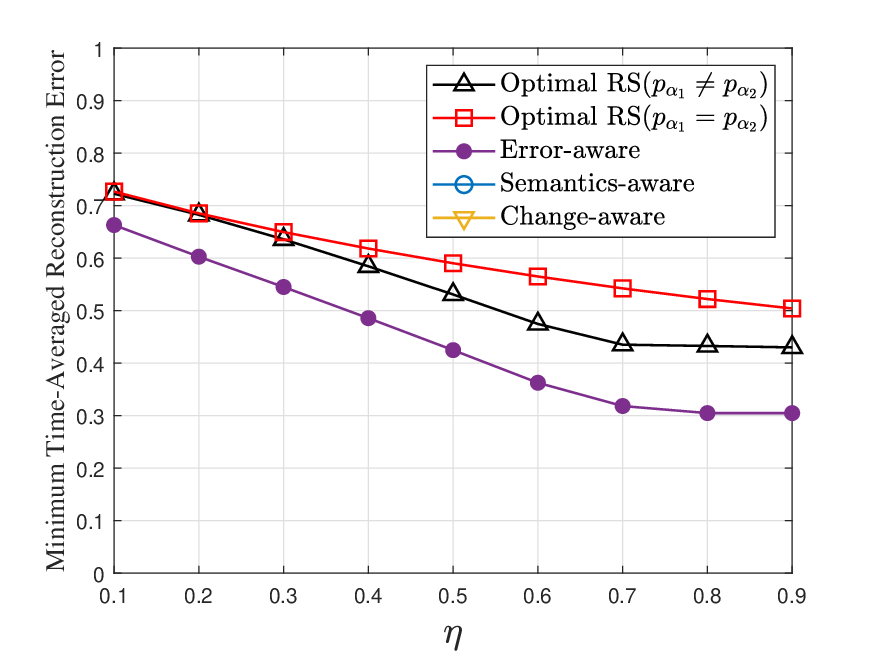}}

\caption{Minimum time-averaged reconstruction error as a function of $\eta$ for selected values of $p=q=0.4$, and success probabilities. For system parameters that violate the cost constraint, the change-aware and semantics-aware policies do not admit feasible solutions; therefore, these cases are excluded from the figures.}
\label{TAvgReconstructionerror_PQFix_ETAVAr2}
\end{figure*}

Figs.~\ref{TAvgSamplingCost_PQVAR1} and \ref{TAvgSamplingCost_PQVAR2} show the time-averaged sampling cost as a function of $p$, for $\eta = 0.8$, and selected values of $q$ and the success probabilities. In these figures, the time-averaged sampling costs for the error-aware and RS policies are computed using the optimal values of $p_{\alpha_1}$, $p_{\alpha_2}$, $q_{\alpha_1}$, and $q_{\alpha_2}$, which minimize the corresponding time-averaged reconstruction error.
The results show that when both $p$ and $q$ are high, indicating strong correlation between $X_1$ and $X_2$, the semantics-aware and change-aware policies result in higher time-averaged sampling costs compared to the other policies. Under these conditions, the sampling cost constraint can be violated, leading to constrained optimization problems with no feasible solutions, as shown in Figs.~\ref{TAvgSamplingCost1_q0.4},~\ref{TAvgSamplingCost2_q0.4},~\ref{TAvgSamplingCost3_q0.4},~\ref{TAvgSamplingCost4_q0.4}. In contrast, under the same values of $p$ and $q$, the optimal error-aware and RS policies adjust their sampling probabilities to ensure that the time-averaged sampling cost remains within the imposed constraint $\eta$. Figs.~\ref{TAvgReconstructionerror_PQFix_ETAVAr1} and \ref{TAvgReconstructionerror_PQFix_ETAVAr2} show the minimum time-averaged reconstruction error as a function of $\eta$, for different levels of statistical correlation between $X_1$ and $X_2$ and various success probabilities. \emph{The results show that the optimal error-aware policy consistently outperforms the other policies, particularly when the time-averaged sampling cost threshold $\eta$ is small and the correlation between $X_1$ and $X_2$ is strong}. Under strong correlation, the semantics-aware and change-aware policies tend to generate more samples to minimize the time-averaged reconstruction error; however, for small $\eta$, this often leads to violations of the sampling cost constraint, resulting in infeasible solutions. In contrast, the optimal error-aware policy utilizes the sampling budget more efficiently by avoiding redundant samples and triggering sampling only when the system is in an erroneous state. This enables a higher effective sampling rate than the optimal RS policy without exceeding the cost constraint. As a result, the error-aware policy achieves a superior trade-off between reconstruction accuracy and resource utilization, making it particularly suitable for resource-constrained environments such as wireless or energy-limited networks.

\section{Conclusion}
We studied the time-averaged reconstruction error in a time-slotted communication system in which two samplers independently observe two correlated processes generated by a common information source and transmit updates over a shared channel. For a broad class of joint sampling and transmission policies, including randomized stationary, error-aware, change-aware, and semantics-aware schemes, we derived closed-form expressions for the stationary distribution and the time-average of the real-time reconstruction error. Building on these analytical characterizations, we formulated and solved a cost-constrained optimization problem to identify the sampling probabilities that minimize the time-averaged reconstruction error for the randomized stationary and error-aware policies. The resulting analysis and numerical evaluations demonstrate that the optimized error-aware policy consistently achieves the lowest reconstruction error while efficiently utilizing the available sampling budget. The performance gains are particularly significant in regimes characterized by strong inter-process correlation and stringent tracking requirements, where accurate real-time monitoring necessitates frequent sampling by both samplers.
\appendix
\subsection{Proof of Lemma {\ref{theorem_PE_RS}}}
\label{Appendix_theorem_PE_RS}
To obtain $\pi_{i,j,m,n}$, we consider a four-dimensional DTMC that describes the joint state of $X_1(t)$ and $X_2(t)$ along with their corresponding error states $E_1(t)$ and $E_2(t)$. In other words, the system state at time $t$ is represented as $\big(X_1(t), X_2(t), E_1(t), E_2(t)\big)$. The one-step transition probabilities $P_{i,j,m,n/i',j',m',n'}=\mathbb{P}\big[X_{1}(t+1)=i',X_{2}(t+1)=j',E_{1}(t+1)=m',E_{2}(t+1)=n'\big|X_{1}(t)=i,X_{2}(t)=j,E_{1}(t)=m,E_{2}(t)=n\big]$ when $X_1(t) = 0$, $E_1(t) = 0$, and $E_2(t) = 0$ under the RS policy are given by:

\begin{align}
    \label{TransProb_RS}
    \scalebox{0.93}{$
    {P}_{0,0,0/0,0,0}$}&\!=\!\scalebox{0.93}{$1-2q,{P}_{0,0,0/0,1,1}=0$},\notag\\
    \scalebox{0.93}{${P}_{0,0,0/1,0,0,0}$}&\!=\!\scalebox{0.93}{$q(1-p_{\alpha_{1}})p_{\alpha_{2}}p_{s_{2/2}}+qp_{\alpha_{1}}p_{\alpha_{2}}p_{s_{2/12}}$},\notag\\
    \scalebox{0.93}{$P_{0,0,0/1,0,0,1}$}&\!=\!\scalebox{0.93}{$qp_{\alpha_{1}}(1\!-\!p_{\alpha_{2}})p_{s_{1/1}}\!+\!qp_{\alpha_{1}}p_{\alpha_{2}}p_{s_{1/12}}(1\!-\!p_{s_{2/12}})$},\notag\\
    \scalebox{0.93}{$P_{0,0,0/1,0,1,1}$}&\!=\!\scalebox{0.93}{$q(1\!-\!p_{\alpha_{1}})(1\!-\!p_{\alpha_{2}}p_{s_{2/2}})\!+\!qp_{\alpha_{1}}(1\!-\!p_{\alpha_{2}})(1\!-\!p_{s_{1/1}})$}\notag\\
    &\!+\!\scalebox{0.93}{$qp_{\alpha_{1}}p_{\alpha_{2}}(1-p_{s_{1/12}})(1-p_{s_{2/12}})$},\notag\\
    \scalebox{0.93}{$P_{0,0,0/1,1,0,0}$}&\!=\!\scalebox{0.93}{$P_{0,0,0/1,0,0,0}$},\scalebox{0.93}{$P_{0,0,0/1,1,0,1}=P_{0,0,0/1,0,0,1}$},\notag\\
    \scalebox{0.93}{$P_{0,0,0/1,1,1,1}$}&\!=\!\scalebox{0.93}{$P_{0,0,0/1,0,1,1}$}.
\end{align}
Similarly, the remaining transition probabilities can be obtained for all possible combinations of $X_{1}(t)$, $X_{2}(t)$, $E_{1}(t)$, and $E_{2}(t)$. Using these transition probabilities, we can then derive the steady-state probabilities $\pi^{\text{RS}}_{i,j,m,n}$ for all $i,j,m,n\in\{0,1\}$ as follows:
\vspace{-0.25cm}
\begin{align}
        \label{pi_RS}
        \scalebox{1}{$
    \pi^{\text{RS}}_{0,0,0}$}&\!\!=\!\! \scalebox{1}{$\frac{p_{\alpha_{1}}p_{s_{1/1}}\big(F+2pG+p_{\alpha_{1}}p_{s_{1/1}}\big)}{(p+2q)\big[\big(2pG\!-\!G\!+\!1\big)p_{\alpha_{1}}p_{s_{1/1}}\!+\!2q\big(pG\!-\!G\!+\!1\big)\big(1\!-\!p_{\alpha_{1}}p_{s_{1/1}}\big)\!\big]}$},\notag\\
        \scalebox{1}{$\pi^{\text{RS}}_{0,1,1}$}&\!\!=\!\!\scalebox{1}{$\frac{2pq\big(1-p_{\alpha_{1}}p_{s_{1/1}}\big)\big(F+pG+p_{\alpha_{1}}p_{s_{1/1}}\big)}{(p+2q)\big[\!\big(2pG\!-\!G\!+\!1\big)p_{\alpha_{1}}p_{s_{1/1}}\!+\!2q\big(pG\!-\!G\!+\!1\big)\big(1\!-\!p_{\alpha_{1}}p_{s_{1/1}}\big)\!\big]}$},\notag\\
        \scalebox{1}{$\pi^{\text{RS}}_{1,0,0,0}$}&\!\!=\!\!\scalebox{1}{$\frac{q\big[p+(1-p)p_{\alpha_{1}}p_{\alpha_{2}}p_{s_{2/12}}+(1-p)(1-p_{\alpha_{1}})p_{\alpha_{2}}p_{s_{2/2}}\big]}{(p+2q)\big[3p\!+\!(1\!-\!3p)p_{\alpha_{1}}p_{\alpha_{2}}p_{s_{2/12}}\!+\!(1-3p)(1\!-\!p_{\alpha_{1}})p_{\alpha_{2}}p_{s_{2/2}}\big]}$},\notag\\
        \scalebox{1}{$\pi^{\text{RS}}_{1,0,0,1}$}&\!\!=\!\!\scalebox{1}{$\frac{pq\big(-Kp^{3}_{\alpha_{1}}+Ip^{2}_{\alpha_{1}}+Dp_{\alpha_{1}}+A\big)}{B(p+2q)\big[\!\big(2pG\!-\!G\!+\!1\big)p_{\alpha_{1}}p_{s_{1/1}}\!\!+\!2q\big(pG\!-\!G\!+\!1\big)\big(1\!-\!p_{\alpha_{1}}p_{s_{1/1}}\big)\!\big]}$},\notag\\
        \scalebox{1.}{$\pi^{\text{RS}}_{1,0,1,1}$}&\!\!=\!\!\scalebox{1}{$\frac{pqGp_{\alpha_{1}}p_{s_{1/1}}}{(p+2q)\big[\!\big(2pG\!-\!G\!+\!1\big)p_{\alpha_{1}}p_{s_{1/1}}\!+\!2q\big(pG\!-\!G\!+\!1\big)\big(1-p_{\alpha_{1}}p_{s_{1/1}}\big)\big]}$},\notag\\
        \pi^{\text{RS}}_{1,1,0,0}&\!=\!\pi^{\text{RS}}_{1,0,0,0}, \pi^{\text{RS}}_{1,1,0,1}=\pi^{\text{RS}}_{1,0,0,1},\pi^{\text{RS}}_{1,1,1,1}=\pi^{\text{RS}}_{1,0,1,1},
    \end{align}
where $A$, $B$, $D$, $I$, $F$, $G$, $H$, $K$, $M$, and $N$ are given by:
\vspace{-0.2cm}
    \begin{align}
    \!\!\!\!\!\scalebox{0.95}{$
        A$}& \!\!= \scalebox{0.95}{$4pqGH + 4qp_{\alpha_{2}} p_{s_{2/2}}  (1-p_{\alpha_{2}}p_{s_{2/2}}) $} ,\label{A_proof}\\
       \!\!\!\!\! \scalebox{0.95}{$B$}&\!\!=\!\scalebox{0.95}{$3p\!+\!(1\!-\!3p)p_{\alpha_{1}}p_{\alpha_{2}}p_{s_{2/12}}\!\!+\!\!(1\!-\!3p)(1\!-\!p_{\alpha_{1}})p_{\alpha_{2}}p_{s_{2/2}}$},\label{B_proof}\\
        \!\!\!\!\!\scalebox{0.95}{$D$}&\!\!=\! \scalebox{0.95}{$p_{\alpha_{2}}  p_{.s_{1/1}}  p_{s_{2/2}} \!\!+\! 4q  p_{s_{1/1}} \!\!-\! 4qp_{\alpha_{2}}p_{s_{1/12}}  (1 \!-\! p_{s_{2/12}})$} \notag\\
\!\!\!\!\!&\!\!\!-\!\scalebox{0.95}{$ 4qp_{\alpha_{2}}p_{s_{2/12}} \!\!\!+\! 
    8qp_{\alpha_{2}}  p_{s_{1/1}}  p_{s_{2/2}} \!\!\!+\!   p(1\!-\!4q)  p_{s_{1/1}}\!G  H $}\notag\\
    \!\!\!\!\!&\!\!\!- 
 \scalebox{0.95}{$p_{\alpha_{2}}^2  p_{s_{2/2}}  \big(p_{s_{1/1}}  p_{s_{2/2}}\! -\! 4qp_{s_{1/1}} \!+\! 4qp_{s_{1/12}}\! -\! 4qp_{s_{1/12}}  p_{s_{2/12}}$}\notag\\
 \!\!\!\!\!&\!\!\!+\scalebox{0.95}{$8q  p_{s_{2/12}} - 4q(2 + p_{s_{1/1}})  p_{s_{2/2}}+4qp_{\alpha_{2}}(p_{s_{2/2}}-p_{s_{1/1}}) \big)$},\label{D_proof}\\
        \!\!\!\!\!\scalebox{0.95}{$I$}&\!\!=\scalebox{0.95}{$4qp_{\alpha_{2}}^2  (p_{s_{2/2}}\!-\!p_{s_{2/12}})  \big(p_{s_{1/12}}  (1 \!-\! p_{s_{2/12}}) \!+\! p_{s_{2/12}} \!-\!
    p_{s_{2/2}}\big)$}\notag\\
    \!\!\!\!\!&\!\!+ \scalebox{0.95}{$(1 - p_{\alpha_{2}})  p_{s_{1/1}}^2  \big(2 - 4q - 
    p_{\alpha_{2}}  p_{s_{2/2}}  (1 - 4q)\big)$} \notag\\
    &\!\!\!- 
   \scalebox{0.95}{$p_{\alpha_{2}}  p_{s_{1/1}}\!(p_{s_{2/2}}\!\!\!-\!p_{s_{2/12}})  \big(1 \!\!-\! 2  p_{\alpha_{2}}  p_{s_{2/2}} \!\!-\! \!
       4q  (2 \!\!-\!\! p_{\alpha_{2}} \!\!-\!\! 2  p_{\alpha_{2}}  p_{s_{2/2}})\big)$} \notag\\
       &\!\!\!-\scalebox{0.95}{$p_{\alpha_{2}}  p_{s_{1/1}} 
    p_{s_{1/12}}  (1 - p_{s_{2/12}})  \big(2 - 4  q - p_{\alpha_{2}}  p_{s_{2/2}}  (1 - 4  q)\big)$},\label{I_proof}\\
        \!\!\!\!\!\scalebox{1}{$F$} &\!\!= \scalebox{0.95}{$p_{\alpha_{2}}p_{s_{2/2}}\!\!-\!p_{\alpha_{1}}p_{\alpha_{2}}\!\big[p_{s_{1/1}}\!\!\!-\!p_{s_{1/12}}(1\!\!-\!\!p_{s_{2/2}})\!-\!p_{s_{2/12}}\!\!+\!p_{s_{2/2}}\big]$},\label{F_proof}\\
            \!\!\!\!\!\scalebox{0.95}{$G$}&\!\!=\scalebox{0.95}{$1-p_{\alpha_{1}}p_{\alpha_{2}}\big[p_{1/12}(1-p_{s_{2/12}})+p_{s_{2/12}}-p_{s_{2/2}}\big]-p_{\alpha_{2}}p_{s_{2/2}}$}\notag\\
            \!\!\!\!\!&\!\!-\scalebox{0.95}{$p_{\alpha_{1}}(1-p_{\alpha_{2}})p_{1/1}$},\label{G_proof}
    \end{align}
 \begin{align}
    \scalebox{0.95}{$H$}&\!\!=\!\!\scalebox{0.95}{$1+p_{\alpha_{1}}p_{\alpha_{2}}(p_{s_{2/2}}-p_{s_{2/12}})-p_{\alpha_{2}}p_{s_{2/2}}$},\label{H_proof}\\
    \!\!\scalebox{0.93}{$K$}&\!\!=\!\!\scalebox{0.95}{$-p_{\alpha_{2}}  p_{s_{1/1}}\!  (p_{s_{2/2}} \!\!-\!\! p_{s_{2/12}})  \big[(1 \!-\! p_{\alpha_{2}})  p_{s_{1/1}} \!\!\!+\! p_{\alpha_{2}}p_{s_{1/12}}  (1 \!\!-\!\! p_{s_{2/12}})$} \notag\\
    &+ \scalebox{0.95}{$p_{\alpha_{2}}p_{s_{2/12}} - p_{\alpha_{2}}p_{s_{2/2}}\big]  (1 - 4  q)$}.\label{K_proof}
    \end{align}
Now, the time-averaged reconstruction error for the RS policy is given by:
    \begin{align}
       \label{PE_RS1}
            &\scalebox{0.98}{$P^{\text{RS}}_{E}$}=\scalebox{0.98}{$\pi^{\text{RS}}_{0,1,1}+\pi^{\text{RS}}_{1,0,0,1}+\pi^{\text{RS}}_{1,0,1,1}+\pi^{\text{RS}}_{1,1,0,1}+\pi^{\text{RS}}_{1,1,1,1}$}.
    \end{align}
Using \eqref{pi_RS}, \eqref{PE_RS1} can be calculated as follows:
    \begin{align}
        \label{PE_RS}
            &\scalebox{1}{$P^{\text{RS}}_{E}$} \notag\\
            &\!=\!\scalebox{1}{$1\!-\!\frac{p_{\alpha_{1}}p_{s_{1/1}}\big(F+2pG+p_{\alpha_{1}}p_{s_{1/1}}\big)}{(p+2q)\big[\big(2pG-G+1\big)p_{\alpha_{1}}p_{s_{1/1}}+2q\big(pG-G+1\big)\big(1-p_{\alpha_{1}}p_{s_{1/1}}\!\big)\!\big]}$}\notag\\
            &-\scalebox{1}{$\frac{2q\big[p+(1-p)p_{\alpha_{1}}p_{\alpha_{2}}p_{s_{2/12}}+(1-p)(1-p_{\alpha_{1}})p_{\alpha_{2}}p_{s_{2/2}}\big]}{(p+2q)\big[3p+(1-3p)p_{\alpha_{1}}p_{\alpha_{2}}p_{s_{2/12}}+(1-3p)(1-p_{\alpha_{1}})p_{\alpha_{2}}p_{s_{2/2}}\big]}$},
        \end{align}
where $F$ and $G$ are as defined in \eqref{F_proof} and \eqref{G_proof}. Similarly, under the error-aware policy, the steady-state probabilities $\pi^{\text{EA}}_{i,j,m,n}$ for all $i,j,m,n\in\{0,1\}$ are given by \eqref{pi_EA} and \eqref{Coeff_EA}. Using \eqref{pi_EA}, the time-averaged reconstruction error under the error-aware policy is given by:
\begin{figure*}[t]
\begin{align}
\label{pi_EA}
\pi^{\text{EA}}_{0,0,0} &= \frac{1}{Z_{1}}\Big[q_{\alpha_{1}}p_{s_{1/1}}\big(F'+2pG'+q_{\alpha_{1}}p_{s_{1/1}}\big)\Big],
\pi^{\text{EA}}_{0,1,1} = \frac{1}{Z_{1}}\Big[2pq\big(1-q_{\alpha_{1}}p_{s_{1/1}}\big)\big(F'+pG'+q_{\alpha_{1}}p_{s_{1/1}}\big)\Big], \notag\\
\pi^{\text{EA}}_{1,0,0,0} &= \frac{1}{Z_{2}}\Big[qN + qq_{\alpha_{1}}p_{s_{2/2}}(1-G')\big(2q + (1-2q)q_{\alpha_{1}}p_{s_{1/1}}\big) + pqMq_{\alpha_{1}}p_{s_{1/1}}\Big],\notag\\
\pi^{\text{EA}}_{1,0,0,1} &= \frac{pq}{Z_{2}}\Big[(pG'+J)q_{\alpha_{1}}p_{s_{1/1}} + 4qW(1-(1-p)G')\Big],
\pi^{\text{EA}}_{1,0,1,1}= \frac{pqG'q_{\alpha_{1}}p_{s_{1/1}}}{Z_{1}},\notag\\ 
\pi^{\text{EA}}_{1,1,0,0} &= \pi^{\text{EA}}_{1,0,0,0}, \quad
\pi^{\text{EA}}_{1,1,0,1} = \pi^{\text{EA}}_{1,0,0,1}, \quad
\pi^{\text{EA}}_{1,1,1,1} = \pi^{\text{EA}}_{1,0,1,1},\\
\label{Coeff_EA}
F' &= q_{\alpha_{2}}p_{s_{2/2}} - q_{\alpha_{1}}q_{\alpha_{2}}\big[p_{s_{1/1}} - p_{s_{1/12}}(1 - p_{s_{2/2}}) - p_{s_{2/12}} + p_{s_{2/2}}\big],\notag\\
G' &= 1 - q_{\alpha_{1}}q_{\alpha_{2}}\big[p_{1/12}(1-p_{s_{2/12}}) + p_{s_{2/12}} - p_{s_{2/2}}\big] - q_{\alpha_{1}}(1-q_{\alpha_{2}})p_{1/1} - q_{\alpha_{2}}p_{s_{2/2}}, \notag\\
Z_{1} &= (p+2q)\Big[(2pG' - G' + 1)q_{\alpha_{1}}p_{s_{1/1}} + 2q(pG' - G' + 1)(1 - q_{\alpha_{1}}p_{s_{1/1}})\Big], \notag\\
Z_{2} &= (p+2q)\Big[(2pG' - G' + 1)q_{\alpha_{1}}p_{s_{1/1}} + 2q(pG' - G' + 1)(1 - q_{\alpha_{1}}p_{s_{1/1}})\Big] \big(3p + (1-3p)q_{\alpha_{2}}p_{s_{2/2}}\big), \notag\\
M &= 3 q_{\alpha_{2}} p_{s_{2/2}} (1 - q_{\alpha_{2}} p_{s_{2/2}}) + q_{\alpha_{1}} q_{\alpha_{2}} p_{s_{1/12}} (1 - p_{s_{2/12}}) (1 - 3 q_{\alpha_{2}} p_{s_{2/2}}) + q_{\alpha_{1}}p_{s_{1/1}} (1 - q_{\alpha_{2}})(1 - 3 q_{\alpha_{2}} p_{s_{2/2}}) \notag\\
&+ q_{\alpha_{1}}q_{\alpha_{2}} (p_{s_{2/2}} - p_{s_{2/12}})(-2 + 3 q_{\alpha_{2}} p_{s_{2/2}}), \notag\\
N &= 4 p q q_{\alpha_{2}} p_{s_{2/2}} (1 - q_{\alpha_{1}} p_{s_{1/1}})(1 - q_{\alpha_{2}} p_{s_{2/2}}) + 2 p^2 G' (1 - q_{\alpha_{2}} p_{s_{2/2}}) (q + (1-q)q_{\alpha_{1}}p_{s_{1/1}}) \notag\\&+ 2 p q q_{\alpha_{1}} (1 - q_{\alpha_{1}} p_{s_{1/1}}) (1 - 2 q_{\alpha_{2}} p_{s_{2/2}}) 
\big[(1 - q_{\alpha_{2}}) p_{s_{1/1}} - q_{\alpha_{2}} (p_{s_{1/12}} (1 - p_{s_{2/12}}) + p_{s_{2/12}} - p_{s_{2/2}})\big], \notag\\
J &= 2 q_{\alpha_{1}} (1 \!-\! q_{\alpha_{2}}) p_{s_{1/1}}\! +\! q_{\alpha_{1}} q_{\alpha_{2}} (2 p_{s_{1/12}} (1 \!-\! p_{s_{2/12}}) \!+\! p_{s_{2/12}}) \!+\! q_{\alpha_{2}} p_{s_{2/2}}(1\!-\!pG') \!-\! q_{\alpha_{1}}q_{\alpha_{2}} p_{s_{2/2}} (1 \!+\! p_{s_{1/1}}\!-\! q_{\alpha_{2}} p_{s_{1/1}}) \notag\\
&- q_{\alpha_{1}}q_{\alpha_{2}}^2 p_{s_{2/2}}(p_{s_{1/12}} + p_{s_{2/12}} - p_{s_{1/12}} p_{s_{2/12}}) - q_{\alpha_{2}}^2 p^2_{s_{2/2}}(1 - q_{\alpha_{1}}), \notag\\
W &= (1 - q_{\alpha_{1}}p_{s_{1/1}})(1 - q_{\alpha_{2}}p_{s_{2/2}}).
\end{align}
\hrule height 0.4pt
\end{figure*}
\vspace{-0.2cm}
        \begin{align}
            \label{PE_EA_proof}
            \scalebox{0.98}{$
            P^{\text{EA}}_{E}
            $}&\!=\scalebox{0.98}{$\pi^{\text{EA}}_{0,1,1}+\pi^{\text{EA}}_{1,0,0,1}+\pi^{\text{EA}}_{1,0,1,1}+\pi^{\text{EA}}_{1,1,0,1}+\pi^{\text{EA}}_{1,1,1,1}$} \notag\\\!
            &=\scalebox{0.98}{$
            1\!-\!\frac{1}{Z_{1}}\big[q_{\alpha_{1}}p_{s_{1/1}}\big(F'\!\!+\!\!2pG'\!\!+\!\!q_{\alpha_{1}}p_{s_{1/1}}\big)\big]\!\!-\!\!\frac{2pqMq_{\alpha_{1}}p_{s_{1/1}}}{Z_{2}}$}
            \notag\\
            &\!\!-\!\!
            \scalebox{0.98}{$
            \frac{2}{Z_{2}}\big[qN+qq_{\alpha_{1}}p_{s_{2/2}}(1-G')\big(2q+(1-\!2q)q_{\alpha_{1}}p_{s_{1/1}}\big)\big] $}.
        \end{align}
Furthermore, under the semantics-aware policy, the steady-state probabilities $\pi^{\text{SA}}_{i,j,m,n}$ for all $i,j,m,n\in\{0,1\}$ can be written as follows:
\begin{align}
\!\!\scalebox{0.88}{$\pi^{\text{SA}}_{0,0,0}$}&\!\!=\!\!\scalebox{0.88}{$\frac{pp_{s_{1/1}}}{L_{1}}\!\Big[p_{s_{2/12}}+(1-p_{s_{2/12}})\big(2p+(1-2p)p_{s_{1/12}}\big)\!\Big]$},\label{pi_SA1}\\
\!\!\scalebox{0.88}{$\pi^{\text{SA}}_{0,1,1}$}&\!\!=\!\scalebox{0.88}{$\frac{2pq}{L_{1}}\big(1\!-\!p_{s_{1/1}}\big)\Big[p_{s_{2/12}}\!+\!(1\!-\!p_{s_{2/12}})\big(2p\!+\!(1\!-\!2p)p_{s_{1/12}}\big)\!\Big]$},\label{pi_SA2}
    \end{align}
\begin{align}
\!\!\scalebox{0.88}{$\pi^{\text{SA}}_{1,0,0,0}$}&\!\!=\!\scalebox{0.88}{$\frac{1}{L_{2}}\!\Big[2qp^{2}(1\!-\!p_{s_{1/12}})(1\!-\!p_{s_{2/12}})(1\!-\!p_{s_{2/2}})\big(q\!+\!(1\!-\!q)p_{s_{1/1}}\big)$}\notag\\
&\!\!+\scalebox{0.88}{$q\big(p_{s_{1/12}}  (1 - p_{s_{2/12}}) + p_{s_{2/12}}\big)\big(2q+(1-2q)p_{s_{1/1}}\big)p_{s_{2/2}}$}\notag\\
&\!\!+\scalebox{0.88}{$2pq^{2}p_{s_{2/2}}\!+\!2pq^{2}(1-2p_{s_{2/2}})\big(p_{s_{2/12}}\!+\!p_{s_{1/12}}(1-p_{s_{2/12}})\big)$}\notag\\
&\!\!+\scalebox{0.88}{$pqp_{s_{1/1}}\!\big(p_{s_{2/2}}\!-\!2qp_{s_{2/2}}\!+\!(2\!-\!3p_{s_{2/2}}\!-\!2q\!+\!4qp_{s_{2/2}})p_{s_{2/12}}\big)$}\notag\\
&\!\!+\scalebox{0.88}{$pqp_{s_{1/1}}p_{s_{1/12}}(1-p_{s_{2/12}})\big(1-2q-(3-4q)p_{s_{2/2}}\big)\Big]$},\label{pi_SA3}\\
\!\!\scalebox{0.88}{$\pi^{\text{SA}}_{1,0,0,1}$}&\!\!=\!\scalebox{0.88}{$\frac{1}{L_{2}}\Big[4pq^{2}\big(p_{s_{2/12}}+(1-p_{s_{2/12}})p_{s_{1/12}}\big)(1-p_{s_{2/2}})$}\notag\\
&\!\!+\scalebox{0.88}{$qp^{2}(1\!-\!p_{s_{1/12}})(1\!-\!p_{s_{2/12}})(1\!-\!p_{s_{2/2}})\big(4q\!+\!(1\!-\!4q)p_{s_{1/1}}\big)$}\notag\\
&\!\!+\scalebox{0.88}{$pqp_{s_{1/1}}p_{s_{1/12}}(1-p_{s_{2/12}})\big(2-4q-(1-4q)p_{s_{2/2}}\big)$}\notag\\
&\!\!+\scalebox{0.88}{$pqp_{s_{1/1}}p_{s_{2/12}}(1-p_{s_{2/2}})(1-4q)\Big]$},\label{pi_SA4}\\
\!\!\scalebox{0.88}{$\pi^{\text{SA}}_{1,0,1,1}$}&\!\!=\!\scalebox{0.88}{$\frac{1}{L_{1}}\Big[pqp_{s_{1/1}}(1-p_{s_{1/12}})(1-p_{s_{2/12}})\Big]$},\label{pi_SA5}\\
\pi^{\text{SA}}_{1,1,0,0}&\!\!=\!\pi^{\text{SA}}_{1,0,0,0},\pi^{\text{SA}}_{1,1,0,1}=\pi^{\text{SA}}_{1,0,0,1},\pi^{\text{SA}}_{1,1,1,1}=\pi^{\text{SA}}_{1,0,1,1}\label{pi_SA6},
    \end{align}
where $L_{1}$ and $L_{2}$ in \eqref{pi_SA1}--\eqref{pi_SA6} are given by:
\vspace{-0.2cm}
     \begin{align}
        \label{T1T2}
        \scalebox{1}{$
        L_{1}$}&=\scalebox{0.9}{$(p + 2 q)  \Big[2p  (1 - p_{s_{1/12}})  (1 - p_{s_{2/12}})  \big(q + 
      (1 - q)p_{s_{1/1}}  \big)$} \notag\\
      &+ \scalebox{0.9}{$\big(p_{s_{1/12}}  (1 - p_{s_{2/12}}) + p_{s_{2/12}}\big)  \big(2  q + 
      (1 - 2q)p_{s_{1/1}}  \big)\Big]$},\notag\\
        \scalebox{0.9}{$L_{2}$}&=\scalebox{0.9}{$\big(3p-3pp_{s_{2/2}}+p_{s_{2/2}}\big)L_{1}$}.
    \end{align}
Using \eqref{pi_SA1}--\eqref{pi_SA6}, the time-averaged reconstruction error for the semantics-aware policy is given by \eqref{PE_SA_lemma_proof}.
\begin{figure*}[t]
\centering
\begin{align}
\label{PE_SA_lemma_proof}
P^{\text{SA}}_{E}
&=
\frac{1}{L_{2}}\Big[
6 q p^{3}(1-p_{s_{1/1}})(1-p_{s_{1/12}})(1-p_{s_{2/12}})(1-p_{s_{2/2}})
+ 2 p q p_{s_{2/12}} 
+ 2 p q p_{s_{1/12}} (1-p_{s_{2/12}})\big(4q+(1-4q)p_{s_{2/2}}\big)\notag\\
&
-8 p q p_{s_{1/1}} p_{s_{2/2}} 
+ 2 p q p_{s_{1/1}} p_{s_{2/12}} (1-3p_{s_{2/2}}\!-\!4q\!+\!4q p_{s_{2/2}})
\!+\!  2 p q p_{s_{1/1}} p_{s_{1/12}} (1-p_{s_{2/12}})
\big(2-4q-(3-4q)p_{s_{2/2}}\big)
 \notag\\
&
+8 p q p_{s_{1/1}} \!+\! 6 p q p_{s_{1/12}} 
\!- 14 p q p_{s_{1/1}} p_{s_{2/12}}
- 6 p q p_{s_{1/12}} p_{s_{2/12}}
+ 14 p q p_{s_{1/1}} p_{s_{1/12}} p_{s_{2/12}} 
+ 2 p q p_{s_{2/2}}
- 8 p q p_{s_{1/12}} p_{s_{2/2}}
\notag\\
&
- 8 p q p_{s_{2/12}} p_{s_{2/2}}
+ 16 p q p_{s_{1/1}} p_{s_{2/12}} p_{s_{2/2}}
+ 8 p q p_{s_{1/12}} p_{s_{2/12}} p_{s_{2/2}} 
+ 8 p q^{2}(1-p_{s_{1/1}})(1-p_{s_{1/12}})
(1-p_{s_{2/12}})(1-p_{s_{2/2}})\notag\\
&
+ 6 p q p_{s_{2/12}}- 14 p q p_{s_{1/1}} p_{s_{1/12}} 
+ 16 p q p_{s_{1/1}} p_{s_{1/12}} p_{s_{2/2}} (1-p_{s_{2/12}})
\Big],\\
\label{T2_proof}
        L_{2}&=\Big[2p  (1 - p_{s_{1/12}})  (1 - p_{s_{2/12}})  \big(q + 
      (1 - q)p_{s_{1/1}}  \big)\!+\! \big(p_{s_{1/12}}  (1 - p_{s_{2/12}}) \!+\! p_{s_{2/12}}\big)  \big(2  q + 
      (1 - 2q)p_{s_{1/1}}  \big)\Big]\notag\\
      &\times (p + 2 q)\big(3p-3pp_{s_{2/2}}+p_{s_{2/2}}\big).
\end{align}
\hrule height 0.4pt
\end{figure*}
Similarly, for the change-aware policy, $\pi^{\text{CA}}_{i,j,m,n}$ for all $i,j,m,n\in\{0,1\}$ can be calculated as:
  \begin{align}
        \scalebox{0.95}{$\pi^{\text{CA}}_{0,0,0}$}&\!=\!\scalebox{0.95}{$\frac{2pp_{s_{1/1}}}{Y_{1}},
        \pi^{\text{CA}}_{0,1,1}\!=\!\frac{p(1-p_{s_{1/1}})\big(1+p_{s_{1/12}}(1-p_{s_{2/12}})+p_{s_{2/12}}\big)}{Y_{1}}$},\label{pi_CA1}\\
        \scalebox{0.95}{$\pi^{\text{CA}}_{1,0,0,0}$}&=\scalebox{0.95}{$\frac{q(1+p_{s_{2/12}})}{(p+2q)(3-p_{s_{2/2}})},\pi^{\text{CA}}_{1,0,1,1}=\frac{qp_{s_{1/1}} (1 - p_{s_{1/12}}) (1 - p_{s_{2/12}})}{Y_{1}}$},\label{pi_CA2}
    \end{align}
    \begin{align}
        \scalebox{0.95}{$\pi^{\text{CA}}_{1,0,0,1}$}&\!=\!\scalebox{0.95}{$\frac{q}{Y_{2}}\Big[\big(1\! +\!p_{s_{1/12}} (1 \!-\! p_{s_{2/12}})\! +\! p_{s_{2/12}}\big) (2 \!-\! p_{s_{2/2}}\!-\!p_{s_{2/12}})$}\notag\\
        &\!-\scalebox{0.95}{$p_{s_{1/1}} (1- p_{s_{1/12}}) (1 - p^{2}_{s_{2/12}})\Big]$},\label{pi_CA3}\\
\pi^{\text{CA}}_{1,1,0,0}&=\pi^{\text{CA}}_{1,0,0,0}, \pi^{\text{CA}}_{1,1,0,1}=\pi^{\text{CA}}_{1,0,0,1},\pi^{\text{CA}}_{1,1,1,1}=\pi^{\text{CA}}_{1,0,1,1},\label{pi_CA4}
    \end{align}
where $Y_{1}$ and $Y_{2}$ in \eqref{pi_CA1}--\eqref{pi_CA4} are given by:
\begin{align}
        \label{Y1Y2}
        \scalebox{0.93}{$
        Y_{1}$}&\!=\! \scalebox{0.93}{$(p+2q)\Big[1\!+\!p_{s_{1/12}}\!+\!(1\!-\!p_{s_{1/12}})\big(p_{s_{2/12}}\!+\!(1\!-\!p_{s_{2/12}})p_{s_{1/1}}\big)\Big]$},\notag\\
         \scalebox{0.93}{$Y_{2}$}&\!=\! \scalebox{0.93}{$(3-p_{s_{2/2}})Y_{1}$}.
\end{align}
Using \eqref{pi_CA1}--\eqref{pi_CA4}, the time-averaged reconstruction error for the change-aware policy can be written as follows:
\vspace{-0.2cm}
    \begin{align}
        \label{PE_CA}
P^{\text{CA}}_{E}&\!=\!\pi^{\text{CA}}_{0,1,1}+\pi^{\text{CA}}_{1,0,0,1}+\pi^{\text{CA}}_{1,0,1,1}+\pi^{\text{CA}}_{1,1,0,1}+\pi^{\text{CA}}_{1,1,1,1}\notag\\
        &\!=\!1-\frac{2pp_{s_{1/1}}}{Y_{1}}-\frac{2q(1+p_{s_{2/12}})}{(p+2q)(3-p_{s_{2/2}})},
    \end{align}
where $Y_{1}$ in \eqref{PE_CA} is derived in \eqref{Y1Y2}.
\vspace{-0.3cm}
\subsection{Proof of Lemmas {\ref{theorem_AvgCost_RS}}, and \ref{theorem_AvgCost_EA}}
\label{Appendix_theorem_timeavg}
The time-averaged sampling cost in \eqref{Optimization_prob1_constraint} for the RS policy is given by:
    \begin{align}
 \label{TAvg_RS}
     &\scalebox{0.9}{$\displaystyle{\lim_{T \to \infty}\frac{1}{T}}\mathbb{E}\Bigg[\displaystyle{\sum_{t=1}^{T}}\delta\Big(\mathbbm1\{\alpha_{1}(t)=1\}+\mathbbm1\{\alpha_{2}(t)=1\}\Big) \Bigg]$} \notag\\
     &=\! \scalebox{0.9}{$\delta\mathbb{P}\Big[\alpha_{1}(t)\!=\!1,X_{1}(t)=0\Big]$}\notag\\
     &+\scalebox{0.9}{$\delta\mathbb{P}\Big[\alpha_{1}(t)\!=\!1,\alpha_{2}(t)\!=\!0, X_{1}(t)=1,X_{2}(t)=0\Big]$}\notag\\
     &+\scalebox{0.9}{$\delta\mathbb{P}\Big[\alpha_{1}(t)\!=\!0,\alpha_{2}(t)\!=\!1, X_{1}(t)=1,X_{2}(t)=0\Big]$}\notag\\
     &+\scalebox{0.9}{$2\delta\mathbb{P}\Big[\alpha_{1}(t)\!=\!1,\alpha_{2}(t)\!=\!1, X_{1}(t)=1,X_{2}(t)=0\Big]$}
     \notag\\
     &+\scalebox{0.9}{$\delta\mathbb{P}\Big[\alpha_{1}(t)\!=\!1,\alpha_{2}(t)\!=\!0, X_{1}(t)=1,X_{2}(t)=1\Big]$}\notag\\
     &+\scalebox{0.9}{$\delta\mathbb{P}\Big[\alpha_{1}(t)\!=\!0,\alpha_{2}(t)\!=\!1, X_{1}(t)=1,X_{2}(t)=1\Big]$}\notag\\
     &+\scalebox{0.9}{$2\delta\mathbb{P}\Big[\alpha_{1}(t)\!=\!1,\alpha_{2}(t)\!=\!1, X_{1}(t)=1,X_{2}(t)=1\Big]$},
 \end{align}
 where using the total probability theorem, \eqref{TAvg_RS} can be written as:
 \begin{align}
 \label{TAvg_RS2}
     &\displaystyle    
     {\scalebox{0.9}{$\displaystyle{\lim_{T \to \infty}\frac{1}{T}}\mathbb{E}\Bigg[\displaystyle{\sum_{t=1}^{T}}\delta\Big(\mathbbm1\{\alpha_{1}(t)=1\}+\mathbbm1\{\alpha_{2}(t)=1\}\Big) \Bigg]$}} \notag\\
     &=\! \scalebox{0.85}{$\delta\mathbb{P}\Big[\alpha_{1}\!(t)\!=\!1\Big|X_{1}\!(t)\!=\!0\Big]\mathbb{P}\big[X_{1}\!(t)=0\big]$}\notag\\
     &\!+\!\scalebox{0.85}{$\delta\mathbb{P}\Big[\alpha_{1}\!(t)\!=\!1,\alpha_{2}(t)\!=\!0\Big| X_{1}(t)\!=\!1,X_{2}(t)\!=\!0\Big]
    \mathbb{P}\big[X_{1}\!(t)\!=\!1,X_{2}(t)\!=\!0\big]$}\notag\\
     &\!+\!\scalebox{0.85}{$\delta\mathbb{P}\Big[\alpha_{1}\!(t)\!=\!0,\alpha_{2}(t)\!=\!1\Big| X_{1}(t)\!=\!1,X_{2}(t)\!=\!0\Big]\mathbb{P}\big[X_{1}(t)\!=\!1,X_{2}(t)\!=\!0\big]$}\notag\\
     &\!+\!\scalebox{0.85}{$2\delta\mathbb{P}\Big[\alpha_{1}(t)\!=\!1,\alpha_{2}(t)\!=\!1\Big| X_{1}\!(t)\!=\!1,X_{2}(t)\!=\!0\Big]\mathbb{P}\big[X_{1}\!(t)\!=\!1,X_{2}(t)\!=\!0\big]$}\notag\\
     &\!+\!\scalebox{0.85}{$\delta\mathbb{P}\Big[\alpha_{1}\!(t)\!=\!1,\alpha_{2}(t)\!=\!0\Big| X_{1}\!(t)\!=\!1,X_{2}(t)\!=\!1\Big]
\mathbb{P}\big[X_{1}\!(t)\!=\!1,X_{2}(t)\!=\!1\big]$}\notag\\
     &\!+\!\scalebox{0.85}{$\delta\mathbb{P}\Big[\alpha_{1}\!(t)\!=\!0,\alpha_{2}(t)\!=\!1\Big| X_{1}(t)\!=\!1,X_{2}(t)\!=\!1\Big]\mathbb{P}\big[X_{1}\!(t)\!=\!1,X_{2}(t)\!=\!1\big]$}\notag\\
     &\!+\!\!\scalebox{0.85}{$2\delta\mathbb{P}\Big[\alpha_{1}(t)\!=\!1,\alpha_{2}(t)\!=\!1\Big| X_{1}(t)\!=\!1,X_{2}(t)\!=\!1\Big]\mathbb{P}\big[X_{1}(t)\!=\!1,X_{2}(t)\!=\!1\big]$}\notag\\
     &=\scalebox{0.85}{$\delta p_{\alpha_{1}}\mathbb{P}\big[X_{1}(t)\!=\!0\big]\!\!+\!\!\delta\big[p_{\alpha_{1}}\!+\!p_{\alpha_{2}}\big]\mathbb{P}\big[X_{1}(t)\!=\!1,\!X_{2}(t)\!=\!0\big]$}\notag\\
     &+\scalebox{0.85}{$\delta\big[p_{\alpha_{1}}+p_{\alpha_{2}}\big]\mathrm{Pr}\big[X_{1}(t)=1,X_{2}(t)=1\big]$}.
 \end{align}
Using the correlated information source shown in Fig.~\ref{system_model_fig},  the probabilities $\mathbb{P}\big[X_{1}(t)=0\big]$, $\mathbb{P}\big[X_{1}(t)=1, X_{2}(t)=0\big]$, and $\mathbb{P}\big[X_{1}(t)=1,X_{2}(t)=1\big]$ can be expressed as: 
 \begin{align}
      \label{jointprobabilities}
      &\!\!\!\!\!\scalebox{0.98}{$\mathbb{P}\big[X_{1}(t)\!=\!0\big]$}\!= \scalebox{1}{$\frac{p}{p+2q}$},\notag\\
       &\!\!\!\!\!\scalebox{0.98}{$\mathbb{P}\big[X_{1}(t)\!=\!1,X_{2}(t)\!=\!0\big]\!=\!\mathbb{P}\big[X_{1}(t)\!=\!1,X_{2}(t)\!=\!1\big]$}\!=\!\scalebox{1}{$\frac{q}{p+2q}$}.
  \end{align}
Now, using \eqref{jointprobabilities}, \eqref{TAvg_RS2} can be simplified as follows:
   \begin{align}
 \label{TAvg_RS3}
     &\lim_{T \to \infty}\frac{1}{T}\mathbb{E}\Bigg[\sum_{t=1}^{T}\delta\Big(\mathbbm1\{\alpha_{1}(t)=1\}+\mathbbm1\{\alpha_{2}(t)=1\}\Big) \Bigg]\notag\\
     &=\frac{\delta\big((p+2q)p_{\alpha_{1}}+2qp_{\alpha_{2}}\big)}{p+2q}.
 \end{align}
Using \eqref{pi_EA} and \eqref{Coeff_EA}, for the error-aware policy, the time-averaged sampling cost in \eqref{Optimization_prob1_constraint} can be written as follows:
\vspace{-0.01cm}
 \begin{align}
     \label{TAvg_EA}
     &\!\!\!\!\scalebox{0.95}{$\displaystyle{\lim_{T \to \infty}\frac{1}{T}}\mathbb{E}\Bigg[\displaystyle{\sum_{t=1}^{T}}\delta\Big(\mathbbm1\{\alpha_{1}(t)=1\}+\mathbbm1\{\alpha_{2}(t)=1\}\Big) \Bigg]$} \notag\\
     &=\scalebox{0.85}{$\displaystyle{\frac{2pq\delta}{Z_{2}}}\Big[3  G'  p^2  q_{\alpha_{1}}  (1 - q_{\alpha_{2}}  p_{s_{2/2}})+q_{\alpha_{1}}  q^{2}_{\alpha_{2}}  p_{s_{2/2}}  (2 p_{s_{1/1}} + p_{s_{2/2}})$}\notag\\
     &\!\!+\!\scalebox{0.85}{$2q^{2}_{\alpha_{1}}q_{\alpha_{2}}  (1 \!\!-\!\! q_{\alpha_{2}})  p_{s_{1/1}}^2\!\!\!+\!\!(q_{\alpha_{1}}q_{\alpha_{2}})^{2}  p_{s_{1/1}}\!  \big(2  p_{s_{1/12}}  (1 \!-\! p_{s_{2/12}}) \!+\! p_{s_{2/12}}\big)$}\notag\\
     &\!\!+\scalebox{0.85}{$(q_{\alpha_{1}}q_{\alpha_{2}})^{2}  p_{s_{2/2}}  \big(p_{s_{1/12}}  (1 - p_{s_{2/12}}) + p_{s_{2/12}} - p_{s_{2/2}}\big)$}\notag\\
     &\!\!+\scalebox{0.85}{$2q^{2}_{\alpha_{1}}q_{\alpha_{2}}  (1 - q_{\alpha_{2}})  p_{s_{1/1}}  p_{s_{2/2}}+4qq_{\alpha_{2}}  (1 - G')    (1 - q_{\alpha_{1}}  p_{s_{1/1}})$}\notag\\
     &\!\!+\scalebox{0.85}{$4pq  q_{\alpha_{2}}  (1 - q_{\alpha_{2}}  p_{s_{2/2}}) +4  p  q_{\alpha_{1}}  q_{\alpha_{2}} (p_{s_{1/1}} + p_{s_{2/2}})  (1 - q_{\alpha_{2}}  p_{s_{2/2}})$}\notag\\
     &\!\!-\scalebox{0.85}{$4p  q_{\alpha_{1}}  q^{2}_{\alpha_{2}}  \big(p_{s_{1/12}}  (1 - p_{s_{2/12}}) + p_{s_{2/12}} - p_{s_{2/2}}\big)-8  p  q  q_{\alpha_{1}}  q_{\alpha_{2}}  p_{s_{1/1}}$}\notag\\
     &\!\!-\scalebox{0.85}{$p  q_{\alpha_{1}}^2 q_{\alpha_{2}}  (p_{s_{1/12}}  (1 - p_{s_{2/12}}) + p_{s_{2/12}} - p_{s_{2/2}})  (-3 + 4  q_{\alpha_{2}}  p_{s_{2/2}})$}\notag\\
     &\!\!+\scalebox{0.85}{$4  p  q  q_{\alpha_{1}}  q^{2}_{\alpha_{2}}   p_{s_{1/1}}  (1 +  p_{s_{2/2}}) -4  p (1 - q)  q_{\alpha_{1}}^2  q_{\alpha_{2}}(1 - q_{\alpha_{2}})   p_{s_{1/1}}^2$} \notag\\
     &\!\!+\!\scalebox{0.83}{$6pq_{\alpha_{1}}^{2}p_{s_{1/1}}\!\!\!-\!\!3pq_{\alpha_{1}}^{2}q_{\alpha_{2}}p_{s_{1/1}}\!\!\!-\!\!7pq_{\alpha_{1}}^{2}q_{\alpha_{2}}p_{s_{1/1}}p_{s_{2/2}}\!\!-\!\!p(q_{\alpha_{1}}q_{\alpha_{2}})^{2}p_{s_{1/1}}p_{s_{2/12}}$}\notag\\
     &\!\!-\!\scalebox{0.85}{$p(q_{\alpha_{1}}q_{\alpha_{2}})^{2}p_{s_{1/1}}\big(-5  p_{s_{2/2}} \!+\! 4  p_{s_{1/12}}  (1 \!-\! p_{s_{2/12}})  (1 \!-\! q) \!-\! 4q p_{s_{2/12}} \big)$}\notag\\
     &\!\!-\scalebox{0.85}{$4pq(q_{\alpha_{1}}q_{\alpha_{2}})^{2}p_{s_{1/1}}p_{s_{2/2}}\Big]$},
 \end{align}
 where $Z_{2}$ and $G'$ are given in \eqref{Coeff_EA}. Similarly, for the semantics-aware policy, the time-averaged sampling cost in \eqref{Optimization_prob1_constraint} can be written as follows:
 \vspace{-0.2cm}
\begin{align}
      \label{TAvg_SA2}
     &\!\!\!\!\scalebox{0.87}{$\displaystyle{\lim_{T \to \infty}\frac{1}{T}}\mathbb{E}\Bigg[\displaystyle{\sum_{t=1}^{T}}\delta\Big(\mathbbm1\{\alpha_{1}(t)=1\}+\mathbbm1\{\alpha_{2}(t)=1\}\Big) \Bigg]$} \notag\\    &\!\!\!\!=\scalebox{0.87}{$\displaystyle{\frac{2pq\delta}{L}}\Big[3p^{2}(1-p_{s_{1/12}})(1-p_{s_{2/12}})(1-p_{s_{2/2}})+4pp_{s_{1/1}}p_{s_{1/12}}p_{s_{2/12}}$}\notag\\
     &+\scalebox{0.87}{$\big(p_{s_{2/12}}
     +(1-p_{s_{2/12}})p_{s_{1/12}}\big)(p_{s_{2/2}}+4q)+p_{s_{1/1}}(p_{s_{2/12}}+2p_{s_{2/2}})$}\notag\\
     &+\scalebox{0.87}{$p_{s_{1/1}}\big(\!-4qp_{s_{2/12}}\!+\!2p_{s_{1/12}}(1-2q)(1-p_{s_{2/12}})\big)\!+\!3pp_{s_{2/12}}\!+\!pp_{s_{2/2}}$}\notag\\
     &+\scalebox{0.87}{$4pq-4pp_{s_{2/12}}(q+p_{s_{2/2}})+pp_{s_{1/12}}  (1 - p_{s_{2/12}})  (3 - 4  p_{s_{2/2}} - 4q)$}\notag\\
     &+\scalebox{0.87}{$pp_{s_{1/1}}\big(7-p_{s_{2/12}}-6p_{s_{2/2}}-4q(1-p_{s_{2/12}})-4p_{s_{1/12}}\big)$}\notag\\
&+\scalebox{0.87}{$4pqp_{s_{1/1}}p_{s_{1/12}}(1-p_{s_{2/12}})\Big]$},
 \end{align}
 where  $L=\Big[2p  (1 - p_{s_{1/12}})  (1 - p_{s_{2/12}})  \big(q + 
      (1 - q)p_{s_{1/1}}  \big)+\big(p_{s_{1/12}}  (1 - p_{s_{2/12}}) + p_{s_{2/12}}\big)  \big(2  q + 
      (1 - 2q)p_{s_{1/1}}  \big)\Big]\big(3p-3pp_{s_{2/2}}+p_{s_{2/2}}\big)(p + 2 q)$.
 Furthermore, for the change-aware policy, we can write the time-averaged sampling cost as follows:
\vspace{-0.15cm}
 \begin{align}
     \label{AvgCost_CA}
      &\!\!\!\!\scalebox{0.87}{$\displaystyle{\lim_{T \to \infty}\frac{1}{T}}\mathbb{E}\Bigg[\displaystyle{\sum_{t=1}^{T}}\delta\Big(\mathbbm1\{\alpha_{1}(t)=1\}+\mathbbm1\{\alpha_{2}(t)=1\}\Big) \Bigg] $}\notag\\ 
      &\!\!\!\!\!=\scalebox{0.87}{$\delta\mathbb{P}\big[X_{1}(t)\!\!=\!\!0\big|X_{1}(t\!-\!1)\!\!=\!\!1,\!X_{2}(t\!-\!1)\!\!=\!\!0\big]\mathbb{P}\big[X_{1}(t\!-\!1)\!\!=\!\!1,\!X_{2}(t\!-\!1)\!\!=\!\!0\big]$}\notag\\
      &\!\!\!\!\!+\scalebox{0.87}{$\delta\mathbb{P}\big[X_{1}(t)\!\!=\!\!0\big|X_{1}(t\!-\!1)\!\!=\!\!1,\!X_{2}(t\!-\!1)\!\!=\!\!1\big]\mathbb{P}\big[X_{1}(t\!-\!1)\!\!=\!\!1,\!X_{2}(t\!-\!1)\!\!=\!\!1\big]$}\notag\\
      &\!\!\!\!\!+\scalebox{0.87}{$2\delta\mathbb{P}\big[X_{1}(t)\!\!=\!\!1,X_{2}(t)=0\big|X_{1}(t\!-\!1)\!\!=\!\!0\big]\mathbb{P}\big[X_{1}(t\!-\!1)\!\!=\!\!0\big]$}\notag\\
      &\!\!\!\!\!+\!\!\scalebox{0.87}{$\delta\mathbb{P}\!\big[\!X_{1}\!(t)\!\!=\!\!1,\!X_{2}(t)\!\!=\!\!0\big|\!X_{1}\!(t\!\!-\!\!1)\!\!=\!\!1,\!X_{2}(t\!\!-\!\!1)\!\!=\!\!1\!\big]\!\mathbb{P}\big[\!X_{1}\!(t\!\!-\!\!1)\!\!=\!\!1,\!X_{2}(t\!\!-\!\!1)\!\!=\!\!1\big]$}\notag\\
       &\!\!\!\!\!+\scalebox{0.87}{$2\delta\mathbb{P}\big[X_{1}(t)\!\!=\!\!1,X_{2}(t)=1\big|X_{1}(t\!-\!1)\!\!=\!\!0\big]\mathbb{P}\big[X_{1}(t\!-\!1)\!\!=\!\!0\big]$}\notag\\
      &\!\!\!\!\!+\!\!\scalebox{0.87}{$\delta\mathbb{P}\!\big[\!X_{1}\!(t)\!\!=\!\!1,\!X_{2}(t)\!\!=\!\!1\big|\!X_{1}\!(t\!\!-\!\!1)\!\!=\!\!1,\!X_{2}(t\!\!-\!\!1)\!\!=\!\!0\!\big]\!\mathbb{P}\big[\!X_{1}\!(t\!\!-\!\!1)\!\!=\!\!1,\!X_{2}(t\!\!-\!\!1)\!\!=\!\!0\!\big]$}\notag\\
      &\!\!\!\!\!=\scalebox{0.87}{$2p\delta\mathbb{P}\big[X_{1}(t)=1,X_{2}(t)=0\big]+2p\delta\mathbb{P}\big[X_{1}(t)=1,X_{2}(t)=1\big]$}\notag\\
      &\!\!\!\!+\scalebox{0.87}{$4q\delta\mathbb{P}\big[X_{1}(t)=0\big]$}.
 \end{align}
 Using \eqref{jointprobabilities}, \eqref{AvgCost_CA} can be simplified as:
 \begin{align}
     \label{AvgCost_CA}
      \!\!\!\!\!\scalebox{0.9}{$\displaystyle{\lim_{T \to \infty}\frac{1}{T}}\mathbb{E}\Bigg[\displaystyle{\sum_{t=1}^{T}}\delta\Big(\mathbbm1\{\alpha_{1}(t)\!=\!1\}\!+\!\mathbbm1\{\alpha_{2}(t)\!=\!1\}\Big) \Bigg]\!=\!\displaystyle{\frac{8pq\delta}{p+2q}}$}.
      \end{align}
\bibliographystyle{IEEEtran}
\bibliography{ref}
\end{document}